\documentclass[journal]{aa}

\usepackage{graphicx}
\usepackage{amsmath}
\usepackage{amssymb}
\usepackage{verbatim}
\usepackage{array}
\usepackage{txfonts}
\usepackage{natbib}
\usepackage[switch,pagewise]{lineno}
\usepackage{multirow}
\usepackage{float}
\usepackage{color}
\usepackage{caption}
\usepackage{xspace}  % This handles spaces after commands    
\usepackage[acronym,nomain]{glossaries}

\newcommand{\celltspace}{\rule{0pt}{2.2ex}}

\newcommand{\dunit}{\,cm$^{2}$\,s$^{-1}$}

\newcommand{\punit}{\,erg\,s$^{-1}$}

\newcommand{\edunit}{\,eV\,cm$^{-3}$}

\newcommand{\gev}{\,GeV\xspace}
\newcommand{\tev}{\,TeV\xspace}
\newcommand{\pev}{\,PeV\xspace}
\def\deg{\ensuremath{^\circ}}

\newacronym{cmb}{CMB}{Cosmic Microwave Background}
\newacronym{cta}{CTA}{Cherenkov Telescope Array}
\newacronym{cr}{CR}{cosmic-ray}
\newacronym{crs}{CRs}{cosmic rays}
\newacronym{gc}{GC}{Galactic Center}
\newacronym{gps}{GPS}{Galactic Plane Survey}
\newacronym{he}{HE}{High Energy}
\newacronym{hgps}{HGPS}{H.E.S.S. Galactic plane survey}
\newacronym{iact}{IACT}{Imaging Atmospheric Cherenkov Telescope}
\newacronym{ic}{IC}{inverse Compton}
\newacronym{ism}{ISM}{interstellar medium}
\newacronym{isrf}{ISRF}{interstellar radiation field}
\newacronym{lmc}{LMC}{Large Magellanic Cloud}
\newacronym{mw}{MW}{Milky Way}
\newacronym{pwn}{PWN}{pulsar wind nebula}
\newacronym{pwne}{PWNe}{pulsar wind nebulae}
\newacronym{sdr}{SDR}{suppressed diffusion region}
\newacronym{sn}{SN}{supernova}
\newacronym{sne}{SNe}{supernovae}
\newacronym{snr}{SNR}{supernova remnant}
\newacronym{snrs}{SNRs}{supernova remnants}
\newacronym{uhe}{UHE}{Ultra High Energy}
\newacronym{vhe}{VHE}{Very High Energy}

\begin{document}

% Title, header, abstract
\title{Are pulsar halos rare ?}
\subtitle{Modeling the halos around PSRs J0633+1746 and B0656+14\\ in the light of {\it Fermi}-LAT, HAWC, and AMS-02 observations \\and extrapolating to other nearby pulsars}
\titlerunning{Are pulsar halos rare ?}
\author{Pierrick Martin\inst{\ref{irap}} 
\and Alexandre Marcowith\inst{\ref{lupm}} 
\and Luigi Tibaldo\inst{\ref{irap}}}
\institute{
IRAP, Universit\'e de Toulouse, CNRS, CNES, F-31028 Toulouse, France \label{irap}
\and Laboratoire Univers et Particules de Montpellier (LUPM) Universit\'e Montpellier, CNRS/IN2P3, CC72, \\Place Eug\`ene Bataillon, F-34095 Montpellier Cedex 5, France \label{lupm}
}
\date{Received 4 March 2022 / Accepted 8 June 2022}
\abstract{Extended gamma-ray emission, interpreted as halos formed by the inverse-Compton scattering of ambient photons by electron-positron pairs, is observed toward a number of middle-aged pulsars. The physical origin and actual commonness of the phenomenon in the Galaxy remain unclear. The conditions of pair confinement seem extreme compared to what can be achieved in recent theoretical models.}
{We searched for scenarios minimizing as much as possible the extent and magnitude of diffusion suppression in the halos in J0633+1746 and B0656+14, and explored the implications on the local positron flux if they are applied to all nearby middle-aged pulsars.}
{We used a phenomenological static two-zone diffusion framework and compared its predictions with {\it Fermi}-LAT and HAWC observations of the two halos, and with the local positron flux measured with AMS-02.}
{While strong diffusion suppression by $2-3$ orders of magnitude at $\sim100$\tev is required by the data, it is possible to find solutions with diffusion suppression extents as small as 30\,pc for both objects. 
If all nearby middle-aged pulsars develop such halos, their combined positron flux including the contribution from Geminga would saturate the $\gtrsim100$\gev AMS-02 measurement for injection efficiencies that are much smaller than those inferred for the canonical halos in J0633+1746 and B0656+14, and more generally with the values typical of younger pulsar wind nebulae. Conversely, if positrons from other nearby pulsars are released in the interstellar medium without any confinement around the source, their total positron flux fits into the observed spectrum for the same injection efficiencies of a few tens of percent for all pulsars, from kyr-old objects powering pulsar wind nebulae to 100\,kyr-old objects like J0633+1746 and B0656+14.}
{It seems a simpler scenario to assume that most middle-aged pulsars do not develop halos, although the evidence supporting that depends on the actual properties of the local pulsar population and on the uncertain physics driving the formation and evolution of halos. The occurrence rate of the phenomenon could be as low as $\sim5-10$\%, and the local positron flux in the $\sim0.1-1.0$\tev range would thus be attributed to a few dozens nearby middle-aged pulsars rapidly releasing pairs into the interstellar medium, with a possible contribution over part or most of the range by J0633+1746, and at higher energies by B0656+14.}

\maketitle

% Introduction
\section{Introduction}
\label{intro}

The discovery of extended emission structures around two nearby middle-aged pulsars, with characteristic ages $\gtrsim 100$\,kyr, suggests the possibility of efficient electron-positron pair confinement beyond the pulsar wind nebula stage \citep{Abeysekara:2017b}. \citet{Linden:2017} suggested that such sources could be common, and more instances of these so-called TeV halos or gamma-ray halos have since then been proposed \citep{DiMauro:2020,Albert:2020}. The phenomenon was demonstrated to have a broadband signature, at least in the gamma-ray range from below 10\gev up to above 100\tev \citep{DiMauro:2019a,Aharonian:2021}.

HAWC observations of PSR J0633+1746 (the Geminga pulsar) and PSR B0656+14 (the pulsar in the Monogem ring) clearly indicate diffusion suppression by factors $\sim100-500$ at $\sim100$\tev particle energies, with respect to the average conditions in the Galactic plane, over an extent of at least 20-30\,pc \citep{Abeysekara:2017b}. Several subsequent analyses have tried to account for the growing ensemble of observations with phenomenological models featuring strong diffusion suppression over at least 100-120\,pc scales \citep[e.g.][]{DiMauro:2019a,Manconi:2020}. Alternative scenarios involving shallower diffusion suppression or even no diffusion suppression at all were proposed \citep{Recchia:2021} but demonstrated to only be marginally applicable to the case of PSR J0633+1746 and not be applicable at all to some other halos on energetic grounds \citep{Bao:2021}. Therefore, the slow or suppressed diffusion paradigm still is the reference one.

Theoretically, the question remains essentially open as to how exactly efficient confinement in the vicinity of a middle-aged pulsar is achieved -- in which medium, by which physical mechanism and over which extent and duration -- and how the phenomenon extrapolates to a galactic population of objects located in a variety of environments. The real challenge posed by pulsar halos lies in the scales involved: strong diffusion suppression reaching up to almost three orders of magnitude, potentially over very extended regions of 100\,pc or more, for long durations of hundreds of kyr, and at very high energies of hundreds of TeV, typically not those at which most sources output the bulk of their non-thermal energy.

Recent theoretical attempts to account for such conditions can be cast into two families: (i) self-confinement, where the gradient of escaping pairs triggers kinetic instabilities out of which turbulence develops so as to scatter the particles themselves; (ii) external turbulence, where the pair halo happens to develop in a region of the \gls{ism} where, for some reason, standard magnetohydrodynamical turbulence has the required properties for efficient confinement of $\sim100$\tev particles. A third option could actually be that the pulsar halo develops is preexisting turbulence self-generated by cosmic-ray protons when they escaped the parent \gls{snr}, provided these conditions can be maintained for long enough \citep{Schroer:2021,Schroer:2022}

In the scenario of turbulence of fluid origin, the properties of the required turbulence were demonstrated to be consistent with those typical of the \gls{ism}, although at the lower end in terms of turbulence correlation length \citep{LopezCoto:2018}. A possible explanation for such conditions was proposed in \citet{Fang:2019}, in which the halo develops in strong turbulence produced in the wake of the expanding shock wave of the parent supernova remnant. The idea is shown to be a valid explanation for J0633+1746, provided 1-10\% of the kinetic energy of the remnant is transferred to turbulence with an injection scale of 10\,pc. Nevertheless, the calculation of wave transport does not include any term for turbulence damping or a self-consistent treatment of dilution, such that the question of turbulence survival after long times still deserves some attention. In addition, the scenario would predict that turbulence around PSR B0656+14 be more intense because it lies in a younger (110 vs. 342\,kyr, less time for cascading) and smaller (60 vs. 90\,pc, less spatial dilution) remnant, whereas the opposite trend is inferred from HAWC observations, with a few times higher diffusion coefficient.

In the scenario of turbulence of kinetic origin, the first studies concluded that it cannot account for strong diffusion suppression at very late times \citep{Evoli:2018,Fang:2019}, essentially because pulsar spin-down and proper motion result in streaming at late times being too weak to generate sufficient turbulence. In \citet{Mukhopadhyay:2021}, the authors revisit that scenario and show that a corrected rate of non-linear wave damping significantly increases both the duration and the suppression of particle diffusion, allowing one to account for the properties of the Geminga halo. Significant confinement still is produced with additional ion-neutral wave damping terms and a flat instead of hard injection spectrum, but prospects degrade rapidly when particle injection is diluted in three dimensions. These encouraging prospects are, however, based on injection starting right at pulsar birth, thereby maximizing the energy reservoir available for turbulence growth, and not taking into account proper motion, which limits the dilution of the particle stream. It is not clear whether or how often such favorable conditions can be realized in nature, since early injection could be expected to come with high proper motion (causing the pulsar to rapidly escape its parent nebula or remnant).

Overall, several promising scenarios have been proposed and explored, but reproducing the confinement inferred for at least the Geminga halo seem to push theoretical models at their limits and require rather optimistic input parameters. This questions the extrapolation of the Geminga halo to the whole Galactic population of pulsars. The theoretical works referred to above demonstrate that pair halos can naturally be produced around middle-aged pulsars, but maybe not all with the extreme properties inferred for the halo around J0633+1746. 

We explore in this paper to what extent the inferred magnitude and extent of diffusion suppression in the halos around J0633+1746 and B0656+14 can be minimized. In the phenomenological framework of a static two-zone diffusion model, we aim at finding the minimum possible values for these parameters in the light of: (i) observations of a gamma-ray halo surrounding these systems with {\it Fermi}-LAT and HAWC; (ii) measurements of the local positron flux with AMS-02 up to about 1\tev. In doing so, we incorporate the constraint that the non-thermal particles produced in these two sources are expected to be typical of the conditions inferred for \gls{pwne}, in terms of injection spectrum, acceleration efficiency, and possible escape age. We then extrapolate these minimalist scenarios to the population of nearby pulsars without currently detected halos to assess the corresponding contribution to the local positron flux. 

% Halo model and parameters
\section{Halo model and parameters}
\label{model}

% Halo definition
\subsection{Halo definition}
\label{model:def}

The nature and physical state of the medium in which the few observed pair halos developed are particularly relevant to any attempt to account for the efficient confinement of particles in the vicinity of the pulsar. Several scenarios have been proposed over the past few years.

\citet{Giacinti:2020} define halos as overdensities of relativistic pairs in an \gls{ism} essentially unaffected dynamically or energetically by the pulsar, leaving open the question of how efficient confinement is achieved. The latter can actually result from standard magnetohydrodynamical turbulence with the appropriate properties. \citet{LopezCoto:2018} determined from first-principles numerical simulations of particle transport that a small enough turbulence correlation length of  $\sim 1$\,pc and magnetic field strength of order 3-5$\mu$G are needed, both conditions being satisfied in the \gls{ism}, although at the lower end of the usually assumed range for the turbulence correlation length \citep{Haverkorn:2008}.

Alternatively, \citet{Tang:2019} propose that the extended emission around the Geminga pulsar results from pairs trapped in the relic nebula that typically formed when the reverse shock crushed the young \gls{pwn} and shifted it away from the pulsar. In that picture, the halo emission would be powered by aging particles injected some time ago, their confinement arising from the specific magnetic topology of the relic nebula, while freshly accelerated particles would be found in the compact bow-shock nebula surrounding the pulsar.

Another option put forward by \citet{Fang:2019} states that the halo around Geminga develops inside the parent \gls{snr}, the expansion and evolution of which is a source of fluid turbulence that eventually is responsible for the suppressed diffusion close to the pulsar. A key question here is whether such turbulence imparted in the first few kyr of the remnant's evolution can retain beyond 100\,kyr a high enough energy density at the scales relevant for $\sim$10-100\tev particle scattering.

Irrespective of the specific case of J0633+1746 and B0656+14, the above scenarios are all situations likely to be realized in nature. They ought to have, however, different consequences in terms of halo properties and evolution and extrapolation to a galactic population; for instance, the scenario of a halo developing inside the parent remnant requires a large remnant and/or a low pulsar proper motion, conditions that will not be met by all middle-aged pulsars. It is beyond the scope of this work to explore all these alternatives and their combinations, and instead we will restrict ourselves to a physical setup for a halo in the spirit of the picture sketched in \citet{Giacinti:2020}: 
\begin{enumerate}
\item the halo phase starts when the pulsar becomes supersonic in its surrounding medium, either when it exits the \gls{pwn} that initially developed at the center of the parent \gls{snr}, that then becomes a relic nebula, or the \gls{snr} itself; this occurs when the displacement of the pulsar owing to its natal kick exceeds the size of the nebula or remnant, typically after a few tens of kyr \citep[after exit from the initial nebula, the pulsar becomes supersonic and develop a bow-shock morphology in the cooling ejecta at 70\% of the remnant radius; see][]{Bucciantini:2008};
\item the pulsar then feeds a compact bow-shock nebula out of which relativistic pairs can easily escape, almost unaffected by energy losses in the nebula; in practice, the whole compact nebula is not resolved in our model and treated as a point source isotropically injecting particles \citep[in the case of Geminga, the bow-shock nebula has sub-pc scales, much smaller than the few tens of pc halo; see][]{Posselt:2017}; 
\item particles are then free to propagate diffusively in the surrounding medium, either the remnant interior or the \gls{ism} essentially undisturbed except for the possibility of a suppressed diffusion of unspecified origin; in the latter case, the environmental conditions for (radiative) energy losses are therefore typical of the \gls{ism} rather than representative of a relic nebula.
\end{enumerate}
Such a definition provides a clear temporal separation between the classical young \gls{pwn} and the older halo phases, but is inadequate to properly model transitional objects. Systems in a stage following the dispersion of the original nebula by the remnant's reverse shock and its subsequent mixing may constitute a fair number of bright and extended gamma-ray sources as well as provide an important and specific contribution to the positron flux. Their modeling is however quite complex, well beyond the scope of the present work, and our focus is restricted to the late stages of fully developed halos. Our scheme, featuring delayed particle injection into the halo, offers the advantage of alleviating the dependence of the predictions on the uncertain initial stages in the pulsar's spin-down history, when most of the rotational energy is lost.

% Model framework and parameters
\subsection{Model framework and parameters}
\label{model:pars}

\begin{table*}[t]
\centering
\begin{tabular}{| c | c  c |}
\hline
\celltspace Parameter & J0633+1746 & B0656+14 \\
\hline
\celltspace Pulsar age $t_{\rm age}$ (kyr) & 342 & 110 \\
\celltspace Pulsar distance $d_{\rm PSR}$ (pc) & 250 & 288 \\
\celltspace Pulsar spin-down power $L(t_{\rm age})$ (\punit) & $3.26 \times 10^{34}$ & $3.80 \times 10^{34}$ \\
\celltspace Pulsar spin-down time scale $\tau_0$ (kyr) & \multicolumn{2}{c |}{12} \\
\celltspace Pulsar braking index $n$ & \multicolumn{2}{c |}{3} \\
\celltspace Magnetic field $B$ ($\mu$G) & \multicolumn{2}{c |}{3} \\
\celltspace Radiation fields energy temperatures $T_i$ (K) & \multicolumn{2}{c |}{(2.73,20,5000)} \\
\celltspace Radiation fields energy densities $U_i$ (\edunit) & \multicolumn{2}{c |}{(0.26,0.30,0.30)} \\
\celltspace Injection spectrum low-energy index $\alpha_1$ & \multicolumn{2}{c |}{1.5} \\
\celltspace Injection spectrum high-energy index $\alpha_2$ & 2.6 & 2.4 \\
\celltspace Injection spectrum break energy $E_b$ (GeV) & \multicolumn{2}{c |}{$10^2$} \\
\celltspace Injection spectrum cutoff energy $E_c$ (TeV) & \multicolumn{2}{c |}{$10^3$} \\
\celltspace Injection start time $t_{\rm exit}$ (kyr) &  \multicolumn{2}{c |}{60} \\
\celltspace Suppressed diffusion region size $R_{\textrm{SDR}}$ (pc) &  \multicolumn{2}{c |}{50} \\
\celltspace Suppressed diffusion normalization $D_{\textrm{SDR}}$ (\dunit) & $4 \times 10^{27}$ & $1 \times 10^{28}$ \\
\celltspace Interstellar diffusion normalization $D_{\textrm{ISM}}$ (\dunit) &  \multicolumn{2}{c |}{$2 \times 10^{30}$} \\
\celltspace Diffusion rigidity scaling index $\delta_{\textrm{D}}$ &  \multicolumn{2}{c |}{1/3} \\
\hline
\end{tabular}
\caption{Summary of baseline parameters used in the modeling of the halos around PSR J0633+1746 and PSR B0656+14.}
\label{tab:halopars} 
\end{table*}

With the above definition, and in the absence of a complete and self-consistent physical picture, pulsar halos were described in the framework of a phenomenological model for the release and propagation of electron-positron pairs in the medium surrounding the pulsar. The exact nature of that medium is not defined (e.g. undisturbed \gls{ism}, or stellar-wind bubble, or interior of an old \gls{snr}), nor is it specified by which physical mechanism enhanced confinement is achieved (e.g. self-confinement of streaming pairs, or standard turbulence with small coherence length stirred by an unknown past event). 

All these important questions are hidden in the description of particle transport around the system as a two-zone diffusive process in static conditions. Particles diffuse away spherically in a medium characterized by a two-zone concentric structure for diffusion properties, with an outer region typical of the average \gls{ism} of the Galactic plane, and an inner region where diffusion is strongly suppressed. The applicability of this framework implicitly means that particles with the relevant energies scatter on mean free paths much smaller than the $\sim30-50$\,pc halo size. Under the interpretation that particle diffusion results from transport in a turbulent magnetic field, this implies a small enough parsec-scale turbulence correlation length \citep{LopezCoto:2018}. 

Similarly, the exact conditions and processes by which pairs are accelerated and released out of the small bow-shock nebula are not detailed. We simply assume that, after pulsar exits from the nebula or the remnant at a time $t_{\rm exit}$, electron-positron pairs are injected directly into the surrounding medium from a central sub-pc source that in practice is point-like in the model, with a spectrum is typical of those inferred from the broadband spectral modeling of most \gls{pwne}.

The complete formalism for this two-zone individual halo model is presented in \citet{Tang:2019} in the context of Geminga\footnote{We warn that Eq. 22 in \citet{Tang:2019} contains a mistake in the arguments of the {\it erf} and {\it erfc} functions. The correct form of the spatial kernel in the two-zone case can be found in \citet{DiMauro:2019a}.}. We neglected the impact of the pulsar's proper motion on the halo morphology because the effects are most prominent at low GeV energies, where the morphology remains poorly constrained \citep[for the halo around Geminga; see][]{DiMauro:2019a}. So when using the {\it Fermi}-LAT data for model fitting, we considered as a constraint only the total emission spectrum.

The parameters of the model, grouped into the main aspects of the phenomenon, are listed and discussed below, while selected values for baseline model setups are summarized in Table \ref{tab:halopars}:

\textit{Pulsar}: As default pulsar parameters for J0633+1746 and B0656+14, we adopted the values used in \citet{Abeysekara:2017b}. The age, distance, power, and spin-down history of the pulsar are of crucial importance. The distance and true age of the system are key parameters when it comes to predicting the associated local positron flux, as illustrated in the next section, and we will discuss the effect of variations from the commonly adopted values. In the case of age, the default values are the characteristic ages of the pulsars, which may overestimate the true ages, so we investigated the possibility that the latter is significantly smaller than the former. 

The spin-down power and history of the pulsar set the maximum power available for particle injection. The present-day power of the pulsar, $L(t)$, is computed as:
\begin{equation}
\label{eq:psrpow}
L(t) = \frac{L_0}{ \left( 1 + t/\tau_0  \right)^{2} } = 4\pi^2 I_{\textrm{NS}} \left( \frac{\dot{P}}{P^3} \right)
\end{equation}
Its value is determined from the current period $P$ and period derivative $\dot{P}$ and an assumed neutron star inertia $I_{\textrm{NS}} =10^{45}$\,g\,cm$^2$. Unless available from other means (e.g. historical observation of the parent supernova), the age of the pulsar is approximated to the characteristic age $\tau_c$ of the pulsar, which is the sum of the true age $ t_{\rm{age}}$ and initial spin-down time scale $\tau_0$.
\begin{equation}
\label{eq:psrage}
\tau_c = \frac{P}{2\dot{P}}= \tau_0 + t_{\rm{age}}
\end{equation}
The spin-down history sets the total number of particles released since injection into the halo started and is relevant for long-lived particles that accumulate over long durations and eventually radiate e.g. in the {\it Fermi}-LAT band. The past spin-down power evolution is assumed to be governed by rotational energy loss from dipole radiation only, i.e. with a braking index $n=3$ (already folded into the above equations), for an assumed $\tau_0$ (to which we are only marginally sensitive since we consider delayed injection after a few tens of kyr; see below).

\textit{Injection}: According to our definition for a halo, particles are accelerated at the pulsar wind termination shock and can easily escape out of a sub-pc bow-shock nebula. The latter process can however be quite complex, anisotropic, highly energy-dependent, and charge separated \citep{Bucciantini:2020}. In the absence of any solid prescription as to what spectrum should be used for any given system, from observational evidence or numerical investigations, we adopted the commonly accepted particle injection spectrum $Q(E,t)$ for younger \gls{pwne} either still in their parent remnant or in the bow-shock phase: a broken power-law, typically with a harder/softer shape at low/high energies and a break at $E_b$ in the few hundreds of GeV range\footnote{The physical mechanism underlying such a spectrum is not fully clear yet, but several proposed scenarios rely on two electron populations: low-energy relic electrons injected during the early rapid spin-down phase of the pulsar and high-energy wind electrons continuously released downstream of the wind termination shock \citep[see][and reference therein, for an application to the Crab nebula]{Meyer:2010}; as an alternative, low-energy particles could result from turbulent acceleration downstream of the termination shock, while high-energy particles would be produced by shock acceleration \citep[][this second scenario is more consistent with the model framework used in this work]{Bucciantini:2011}. The physical meaning of the break energy depends on the scenario: either low-energy cutoff of the high-energy population reflecting the Lorentz factor of the bulk upstream flow \citep{Kennel:1984}; or high-energy cutoff of the low-energy population reflecting the maximum attainable energy in the turbulent flow \citep{Bucciantini:2011}. In either case, a smooth connection of the two population remains a challenge.}. We note, however, that the relevance of such a spectral shape for much older pulsars or in the context of halos is not solidly demonstrated, and that harder single power-law shapes seem to be a viable alternative for several objects \citep{Sudoh:2021}.
\begin{equation}
\label{eq:injspec}
Q(E,t) = Q_0(t) \times e^{-E/E_c} \times \begin{cases} (E/E_b)^{-\alpha_1} & \mbox{if } E\leq E_b \\
(E/E_b)^{-\alpha_2} & \mbox{if } E > E_b
\end{cases}\,
\end{equation}
The low-energy part of the spectrum is hardly constrained by the data we are comparing to; the power-law index $\alpha_1$ is inferred to lie in the 1.0-2.0 range from radio observations \citep{Bucciantini:2011,Torres:2014}, and we fixed it to a mean value of 1.5. Conversely, the high-energy index $\alpha_2$ does strongly influence the predictions for the observables we are considering. It is constrained to values in the range 2.0-2.8 from spectral fits to various \gls{pwne} \citep{Bucciantini:2011,Torres:2014,Principe:2020,Abramowski:2015}, and we therefore investigated the effect of changing the index within this range. For the break energy $E_b$, we adopted a typical value of 100\gev, and checked that values up to 500\gev have little influence on the qualitative and quantitative conclusions of the study. Last, the accelerated particle spectrum is expected to feature a cutoff at very high energies reflecting the time evolution of the potential drop, hence decreasing from a few PeV down to a few hundreds of TeV over the first few hundreds of kyr of the pulsar \citep{Evoli:2021}. We however adopted a constant spectral cutoff at $E_c=1$\pev since the observables we are comparing to are marginally sensitive to particle energies above a few hundreds of TeV.
 
Particle acceleration is powered by the spin-down luminosity of the pulsar and the injection efficiency $\eta$, i.e. the fraction of the pulsar spin-down power converted into non-thermal pairs, is a free parameter in the model. 
\begin{equation}
\label{eq:injpow}
\eta L(t) = \int_{{\rm{1\gev}}}^{{\rm{1\pev}}} E Q(E,t) dE \\
\end{equation}
The modeling of the broadband emission from \gls{pwne} suggests relatively large values for $\eta$, of the order of a few tens of percent and reaching up to nearly 100\% \citep{Bucciantini:2011,Torres:2014}. We emphasize that the spin-down power is computed under the hypothesis of a typical moment of inertia of $10^{45}$\,g\,cm$^2$, and variations by up to factor 2 are possible depending on the actual mass and internal structure of the neutron star \citep{Worley:2008,Zhao:2016}. In this respect, the 100\% efficiency is no strict limit and injection efficiencies moderately in excess of that value could be considered. It seems quite an extrapolation to consider that high injection efficiencies close to 100\% inferred for kyr-old objects could still hold for the 100\,kyr-old pulsars powering halos. Nevertheless, as we will see below, the injection efficiency required for the two canonical halos around J0633+1746 and B0656+14 are of that order.

Finally, an important parameter is the time when non-thermal particle release into the halo begins; several studies assumed an injection starting at pulsar birth \citep[see][for a discussion of the effects of such a parameter in the context of a population synthesis]{Sudoh:2019}, but this overlooks the \gls{pwn} phase where particles are injected and trapped into an expanding, highly magnetized nebula for a few tens of kyr. Several studies considered that pairs from pulsars are fed into the \gls{ism} when the pulsar exits its parent remnant owing to its natal kick. Estimates for typical exit times are in the 40-60\,kyr range \citep[see e.g.][]{Bykov:2017,Evoli:2021}. In this study, we assumed by default an injection start time of $t_{\rm exit}=60$\,kyr, but we considered variations from 20 to 80\,kyr and assessed their impact on our results.

\textit{Diffusion}: in the two-zone diffusion framework we are using, particles released from a point-like source diffuse away spherically in a medium characterized by a two-zone concentric structure for diffusion properties, with an outer region typical of the average \gls{ism} of the Galactic plane, and an inner region where diffusion is strongly suppressed:
\begin{equation}
\label{eq:diff}
D(E, r) = \left( \frac{E}{100 \rm TeV} \right)^{\delta_{\rm D}} \times \begin{cases}
D_{\rm SDR}, \, 0 < r < R_{\rm SDR},\\
D_{\rm ISM}, \, r \geq R_{\rm SDR},
\end{cases}\,
\end{equation}
The two key parameters of the problem are the radial extent of the suppressed diffusion region, $R_{\rm SDR}$, and the value of the diffusion coefficient therein, $D_{\rm SDR}$, and they will be the main focus of our work. We also discuss below the impact of the diffusion index $\delta_{\rm D}$.

The particle distribution around the pulsar in the two-zone diffusion framework is computed following the formalism detailed in \citet{Tang:2019}:
\begin{align}
&N(r, E) = \int_{{\rm max}[t_{\rm exit},\,t_{\rm exit}-t_{\rm cool}(E_c, E)]}^{t_{\rm age}} \frac{\dot{E}(E_0)}{\dot{E}(E)}Q(E_0,t_0) H(r,E) dt_0 \\
&t_{\rm cool}(E_0,E) = \int_{E}^{E_0} \frac{-dE}{\dot{E}}
\label{eq:soltwozone}
\end{align}
The full expression for the spatial kernel $H(r,E)$ in the two-zone case is given in \citet{Tang:2019} and \citet{DiMauro:2019a} and we do not repeat it here. The integration runs over the full injection history, i.e. over all particles released at time $t_0$ with an energy $E_0$ that decreased as a result of losses to $E$ at time $t_{\rm age}$. 

Recent theoretical efforts to quantify the extent of self-confinement of escaping non-thermal particles in the case of \gls{snrs} or pulsars suggest that diffusion suppression by 1-2 orders of magnitude could be reached over 50-100\,pc distances at most \citep{Malkov:2013,Nava:2016,Nava:2019,Evoli:2018,Schroer:2022}. This is consistent with the levels and extents inferred from observations or modeling of other objects \citep[see][for a recent review]{Tibaldo:2021}. Around individual objects such as \gls{snrs} or pulsars, sustaining this strong confinement over durations above 100\,kyr is a challenge, especially at very high energies \citep[but see][for recent developments in the case of pulsars]{Mukhopadhyay:2021}. 

In the case of particles diffusing inside a very large and possibly relic nebula, a likely maximum extent is suggested by the largest known \gls{pwn}, HESS~J1825-137, that has a radius of 100\,pc \citep{Principe:2020}. The very nature of that object remains unclear and, interestingly, it was recently suggested to be a mixed or transitional object evolving from a classical \gls{pwn} to a halo \citep{Giacinti:2020}. In the case of trapping in the enhanced fluid turbulence that could be expected in the interior of an old and large \gls{snr} or inside a stellar-wind bubble, the maximum size of such objects is of the same order, 50-100\,pc. \gls{snrs} with ages $\lesssim20$\,kyr seem to have radii below 30\,pc \citep{Badenes:2010}, while older objects $\lesssim100$\,kyr can reach up to 100-120\,pc \citep{Fesen:2021}. A reasonable maximum extent for the suppressed diffusion region therefore seems to be 100\,pc, and we will therefore favor values below such a limit, if data allow.

Overall, the results referred to above suggest that diffusion suppression by up two orders of magnitude over an extent of 50-100\,pc at most should be considered a reasonable limit, and we will try to find model setups for J0633+1746 and B0656+14 that comply with such bounds, as much as data allow. As we will show below, however, the magnitude of diffusion suppression seems pretty well constrained by HAWC observations, at least at few hundreds of TeV particle energies and in the two-zone diffusion framework, and the halo around Geminga points to diffusion suppression reaching almost 3 orders of magnitude.

The rigidity dependence of the suppressed diffusion parameter, here approximated as an energy dependence, is another possibly important free parameter that may well differ from standard Kolmogorov or Kraichnan scaling, especially in the case of self-confinement \citep[e.g.][]{Evoli:2018}. This parameter is however strongly degenerate with others: HAWC observations pinpoint the diffusion properties for short-lived freshly injected $\sim100$\tev particles, but the spatial distribution of the much longer-lived $\sim0.1-1$\tev particles contributing to the positron flux or shining in the {\it Fermi}-LAT band can be reproduced by several combinations of diffusion coefficient at this energy, injection spectrum, and time when injection started. So for simplicity, we assumed by default that suppressed diffusion is, like average interstellar diffusion, described by a Kolmogorov scaling $\delta_{\rm D}=1/3$, but we will discuss below the effect of a harder rigidity dependence. For completeness, we emphasize that the exact shape and extent of the transition from suppressed to average interstellar diffusion also matters, but this is second order here and would only add unnecessary complexity \citep{Profumo:2018}. 

The value of the effective diffusion coefficient in the outer region can be considered as relatively well determined from studies of cosmic-ray propagation in the Galaxy. One cannot exclude, however, the presence of inhomogeneities in transport properties throughout the disc, even beyond very localized effects around sources (such as those around \gls{snrs} or maybe at stake around Geminga). This is however poorly constrained at the moment so we stuck to the average value for the Galactic plane, using by default a value consistent with the range in \citet{Trotta:2011} for the normalization and a Kolmogorov scaling for the rigidity dependence.

\textit{Energy losses and radiation}: Synchrotron and inverse-Compton scattering losses drive the propagation range of the $\gtrsim1$\tev particles, which are in a loss-dominated regime. The energy loss rate $\dot{E}$ of particle including the Klein-Nishina regime was approximated following \citet{Moderski:2005}:
\begin{eqnarray}
\label{eq:tcool}
\dot{E} = \dot{E}_{IC} +\dot{E}_{syn} = -\frac{4}{3}c\sigma_T \gamma^2\left[ \sum_i\frac{U_{i}}{(1+4\pi \gamma \epsilon_i)^{3/2}} + \frac{B^2}{8\pi}\right]
\end{eqnarray}
The sum is performed over all components of the radiation field, described as graybodies with energy densities $U_i$ and temperatures $T_i$ (see the values in Table \ref{tab:halopars}), yielding normalized photon energies for the fields $\epsilon_i = 2.8k_BT_i/m_ec^2$, where $k_B$ is the Boltzmann constant, $m_e$ the electron mass, and $c$ the speed of light. By default, we used the same magnetic and radiation fields as in \citet{Abeysekara:2017b}.

For the magnetic field, we used a baseline value of $B=3$\,$\mu$G, typical of the local \gls{ism}, but considered variations by factors of a few, since the medium in which the halos develop may not be representative of the average \gls{ism} (e.g., if it is the interior of the parent remnant). As an alternative for the radiation field, which sets both the magnitude of losses and the seed photon fields for inverse-Compton scattering, we tested the model of \citet{Popescu:2017} for the position of Geminga in the Galaxy and found a limited impact on model predictions. More generally, there are obvious uncertainties in magnetic and radiation fields in the medium surrounding the pulsar, and we investigated the impact of varying the input values, as described in Sect. \ref{res:efffields}.

Inverse-Compton radiation from the halo is computed from the projected particle angular distribution, $P(\theta,E)$, obtained by integrating the particle spatial distributions along the line of sight for the assumed distance $d_{\rm PSR}$ to the pulsar:
\begin{equation}
P(\theta,E) = 2  d_{\rm PSR}^2 \times \int_{0}^{r_{\rm max}}{N \left( (\theta^2 d_{\rm PSR}^2+\ell^2)^{1/2},E \right) d\ell}
\end{equation}
In practice, the particle spatial distribution $N(r, E)$ is computed over a grid of radial positions from 0 to $r_{\rm max} = {\rm max}(100\,{\rm pc},R_{\textrm{SDR}}+20\,{\rm pc}$, with a step of 1\,pc by default\footnote{We tested the effect of grid parameters and found them to be negligible except at the lowest energies when diffusion suppression is very low, much lower than the values handled here (because the particle distribution then becomes very extended within the suppressed diffusion region and particle density is non-negligible outside of it).}. The corresponding inverse-Compton radiation is computed from the particle angular distributions using the Naima package \citep{Zabalza:2015}, in the approximation of isotropic radiation fields\footnote{Inverse-Compton scattering radiation should ideally be computed taking into account the anisotropic nature of the seed photon fields, but the effect was found to be minor and most likely degenerate with other free parameters of the problem \citep{Johannesson:2019}.}

In the above parameters, some are object-specific and will only affect the modeling of J0633+1746 or B0656+14 (e.g. the pulsar age), while others are bound to have an impact on a putative halo population if they are taken as representative (e.g. the suppressed diffusion region properties). A population perspective, built on the (still questionable) assumption that all middle-aged pulsars develop at some stage a halo similar to that of J0633+1746 or B0656+14, bear some potential for constraining some of the parameters of the problem, at least in a statistical sense. We illustrate that point below in Sect. \ref{nearby}, where we extrapolate the model setups for halos like J0633+1746 or B0656+14 to the population of nearby pulsars to assess the corresponding contribution to the local positron flux. A complementary exercise involving the synthesis of an entire Galactic population of halos around mock or real pulsars and the analysis of the resulting source flux distribution in the light of existing measurements in the GeV and TeV range will be presented in subsequent publications.

\begin{figure}[!t]
\begin{center}
\includegraphics[width=0.9\columnwidth]{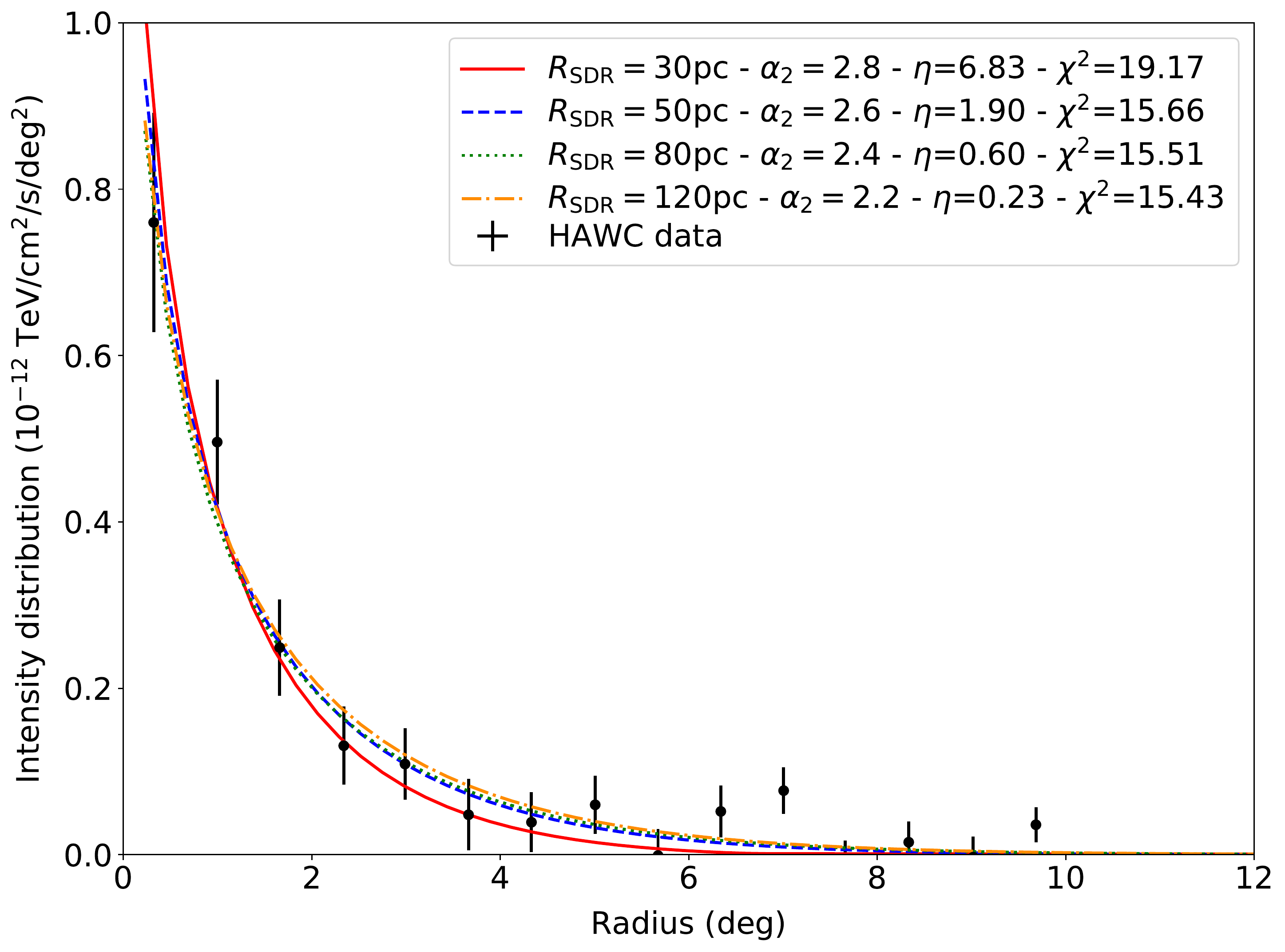}
\includegraphics[width=0.9\columnwidth]{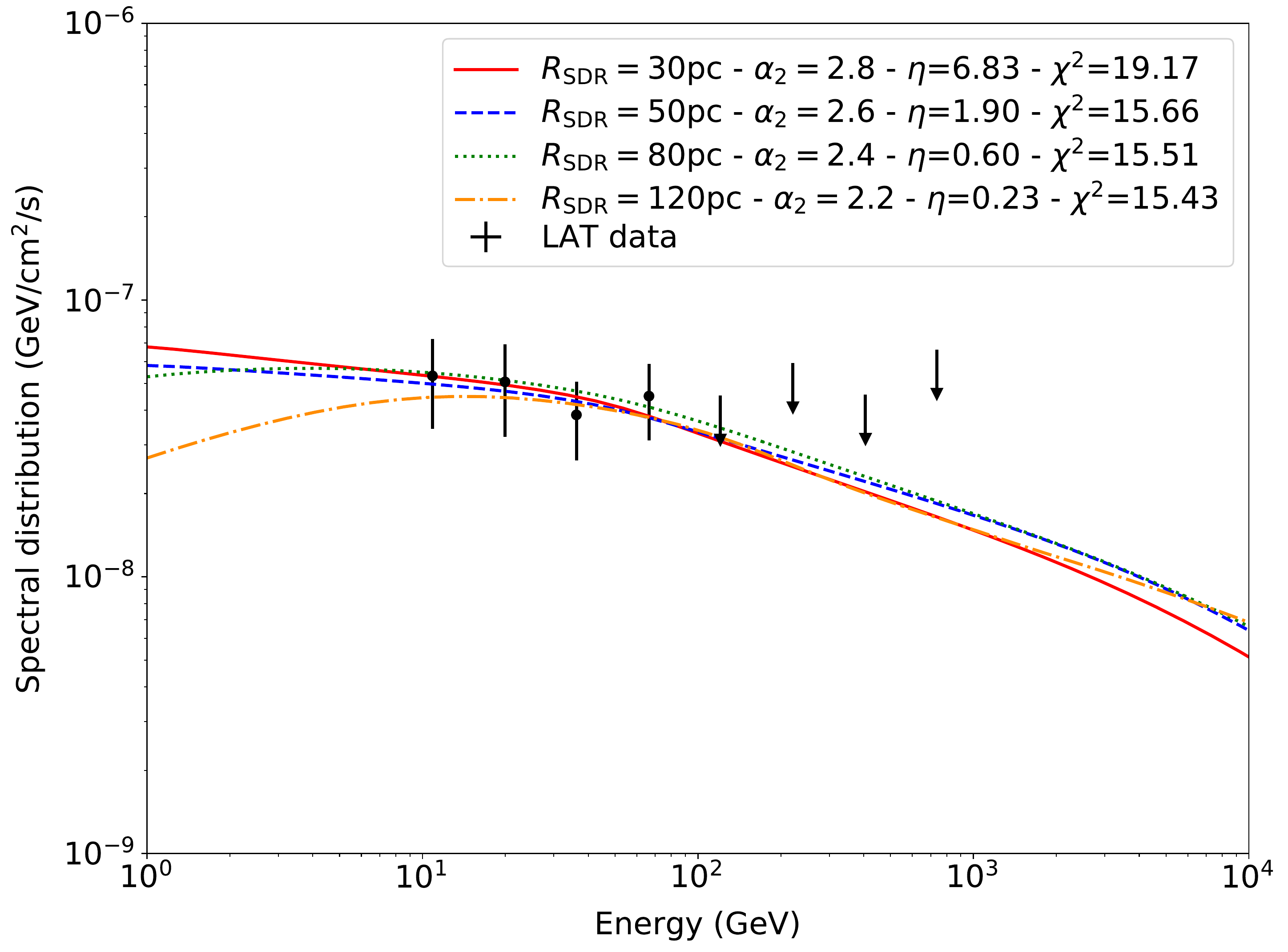}
\includegraphics[width=0.9\columnwidth]{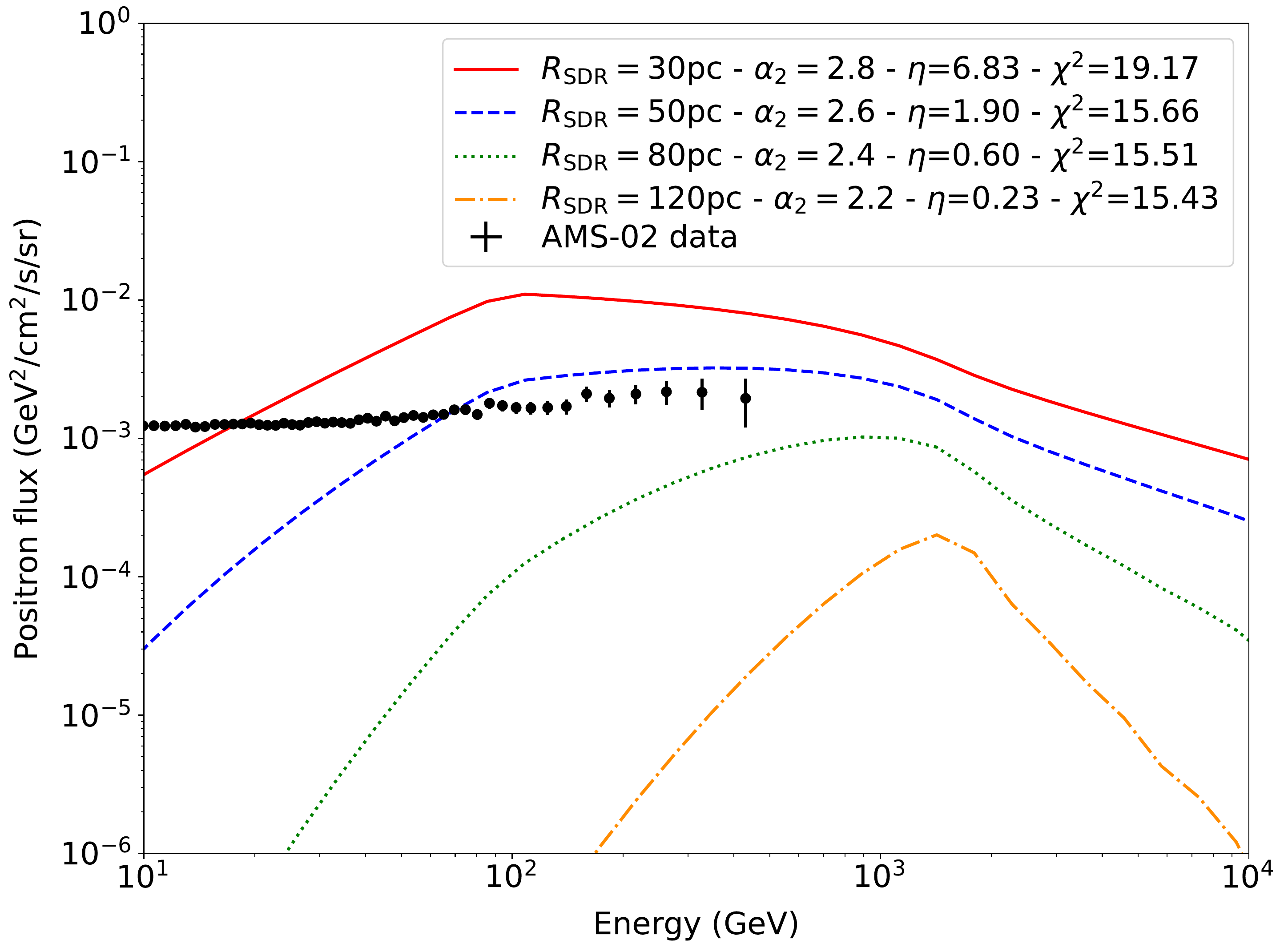}
\caption{Model fits to HAWC+LAT observations of J0633+1746, and the corresponding predicted local positron fluxes, for different sizes of the suppressed diffusion region $R_{\rm SDR}$, and different high-energy injection index $\alpha_2$. All other parameters have their baseline values, in particular the diffusion coefficient $D_{\textrm{SDR}}(100\textrm{TeV}) = 4 \times 10^{27}$\dunit. The set of dashed blue curves corresponds to the baseline model setup. The legend indicates the injection efficiency $\eta$ and the $\chi^2$ of the fit.}
\label{fig:mod:geminga:effect-size}
\end{center}
\end{figure}

\begin{figure}[!t]
\begin{center}
\includegraphics[width=0.9\columnwidth]{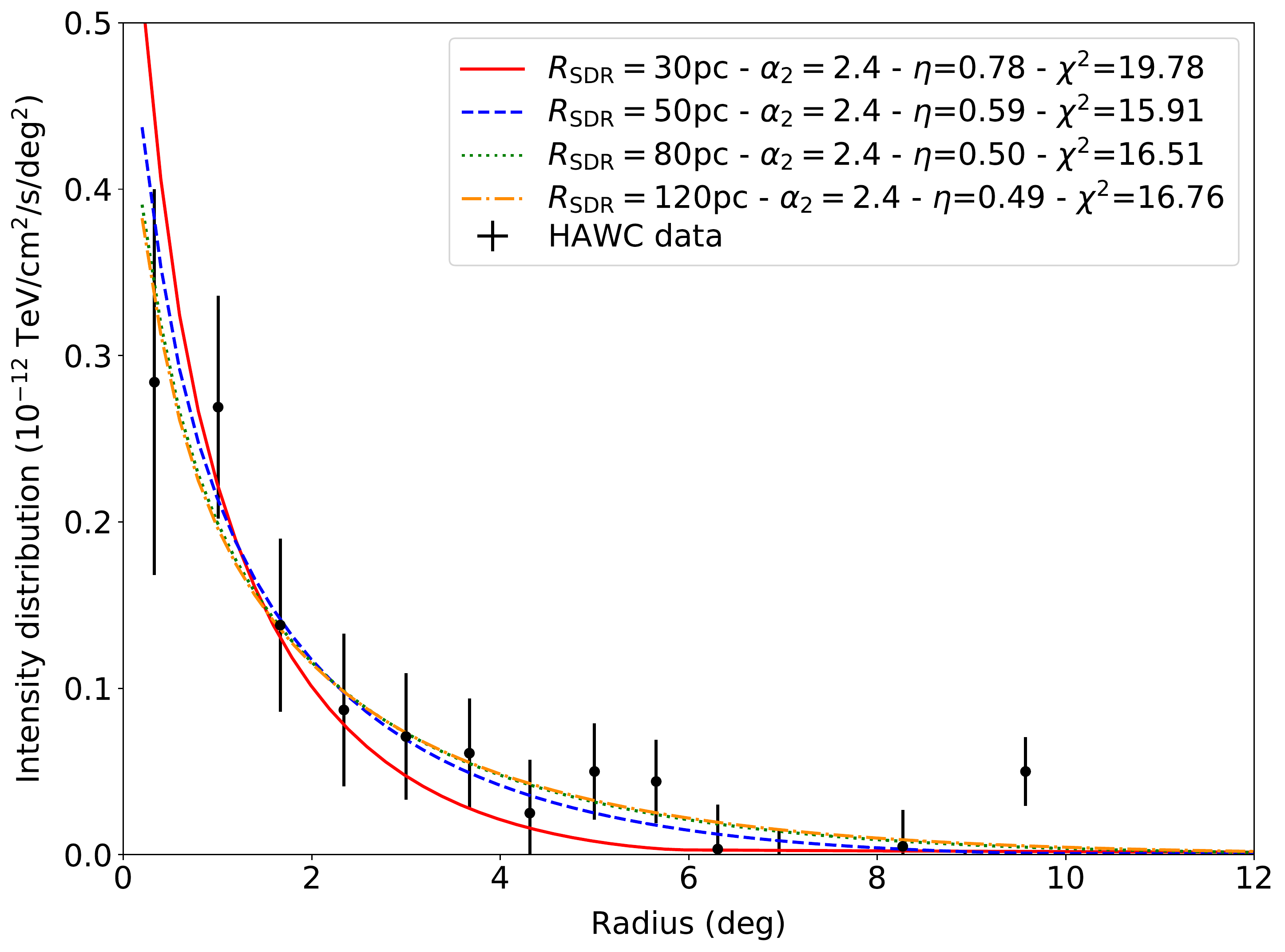}
\includegraphics[width=0.9\columnwidth]{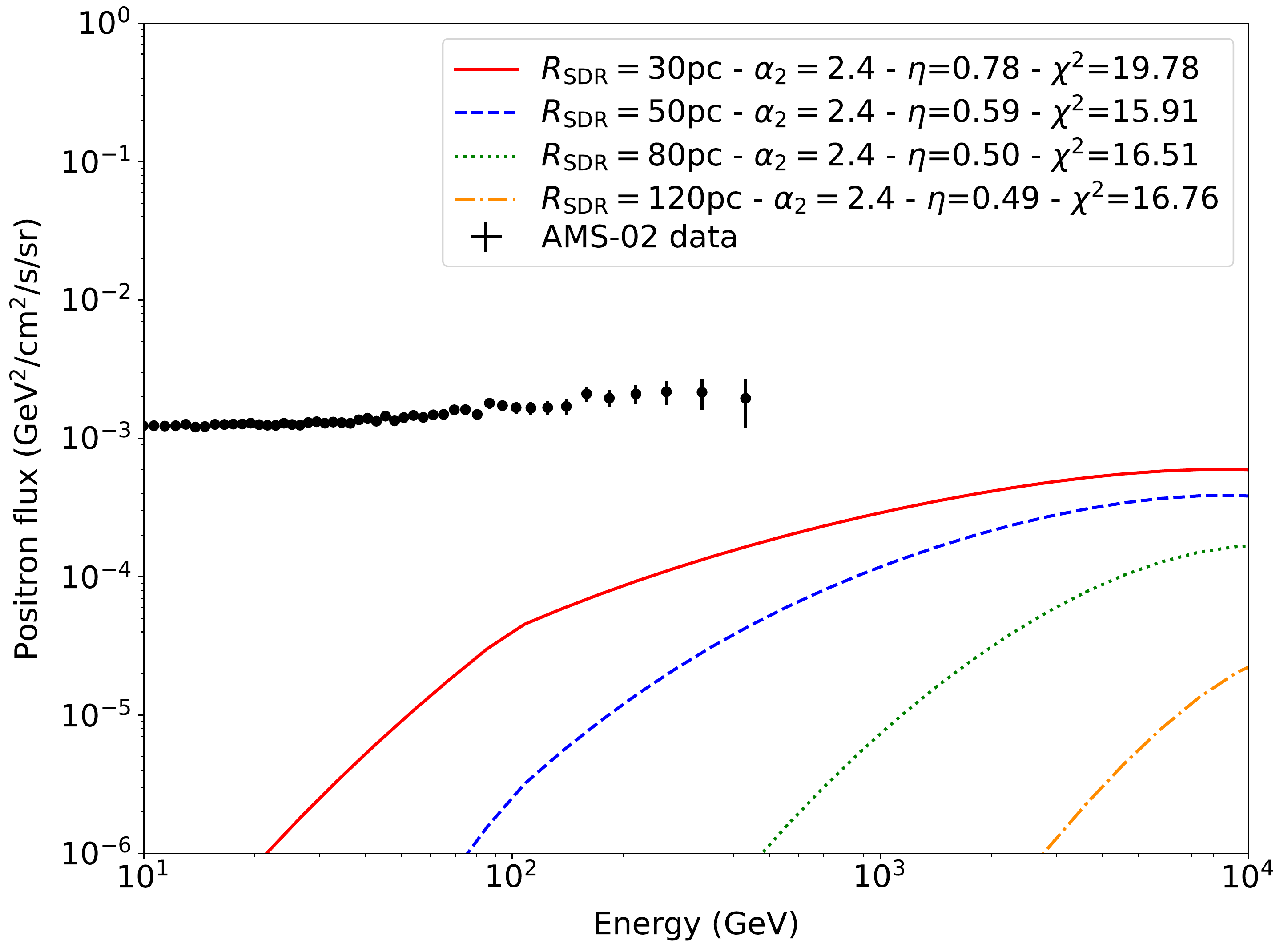}
\caption{Model fits to HAWC observations of B0656+14, and the corresponding predicted local positron fluxes, for different sizes of the suppressed diffusion region $R_{\rm SDR}$. All other parameters have their baseline values, in particular the diffusion coefficient $D_{\textrm{SDR}}(100\textrm{TeV}) = 10^{28}$\dunit. The set of dashed blue curves corresponds to the baseline model setup. The legend indicates the injection efficiency $\eta$ and the $\chi^2$ of the fit.}
\label{fig:mod:monogem:effect-size}
\end{center}
\end{figure}

% Results
\section{Results}
\label{res}

In this section, we start from baseline model setups for the halos around J0633+1746 or B0656+14 and discuss several possibilities for minimizing the extent and magnitude diffusion suppression, both in extent and magnitude, while still being in agreement with the HAWC and {\it Fermi}-LAT gamma-ray observations, and with the local positron flux measured with AMS-02.

% Baseline model setups
\subsection{Baseline model setups}
\label{res:base}

Our baseline model setups for J0633+1746 or B0656+14 are inspired by previous modeling efforts \citep[e.g.][]{Abeysekara:2017b,Tang:2019,DiMauro:2019a,Johannesson:2019} and the expectation that particle injection in the halos is similar to what is inferred for younger \gls{pwne} \citep{Bucciantini:2011,Torres:2014,Zhang:2008}. The corresponding parameter sets are summarized in Table \ref{tab:halopars}. The only remaining free parameter is the injection efficiency, which is determined by a joint fit to the LAT spectrum and HAWC intensity distribution. 

The predictions for these baseline model setups are shown as dashed blue lines in the curve series presented in Figs. \ref{fig:mod:geminga:effect-size} and \ref{fig:mod:monogem:effect-size}. A good match to the gamma-ray observations is obtained in either case. The optimal suppressed diffusion coefficient normalizations are about $4 \times 10^{27}$ and $1 \times 10^{28}$\dunit\ at 100\tev, for J0633+1746 and B0656+14, respectively, in agreement with estimates from \citet{Abeysekara:2017b}. Yet, while the fit for B0656+14 involves an acceptable injection efficiency of about 60\%, that for Geminga requires an excessive value of nearly 200\%. This stems from the fact that matching both the LAT and HAWC data calls for a rather soft injection spectrum, with an index of 2.6. Clearly, the baseline setup is not a viable option for Geminga, and we will show in the next subsections how solutions with acceptable efficiencies can be found.

The suppressed diffusion coefficient at $\sim100$\tev is set by the intensity profiles observed with HAWC, for the assumed distances and ages of the pulsars, while the LAT data set the slope of the injection spectrum above the break energy $E_b$ and the overall acceleration efficiency (for the assumed broken power-law spectral shape). In all presented results, the predicted spectrum in the {\it Fermi}-LAT range is not shown for B0656+14, because the available upper limits in this range were found to be poorly constraining for all model setups that were investigated. This is due to the relatively small age of the pulsar, which leaves little time for the accumulation of low-energy particles after injection start at 60\,kyr, so a relatively weaker flux in the GeV band, such that the {\it Fermi}-LAT upper limit is easily accommodated. 

The bottom panel of Fig. \ref{fig:mod:geminga:effect-size} shows that, in addition to requiring an unphysical injection efficiency, the baseline setup for Geminga yields a positron flux exceeding the AMS-02 measurements above $\sim100$\gev, while the predicted flux in the case of B0656+14 is fully consistent with it. The qualitative differences between J0633+1746 or B0656+14 in positron flux come mostly from the smaller age of the latter (110 vs. 342\,kyr) which implies less accumulation of low-energy particles and less time for diffusion, and as a result a positron flux peaking at $\sim10$\tev, above the current reach of AMS-02.

These results are a direct outcome of using a two-zone diffusion model with a limited extent for the suppressed diffusion region. In the extreme case of an infinitely large confinement region, in practice a one-zone model, the local positron flux is heavily suppressed \citep{Abeysekara:2017b}. Finite confinement regions with very large sizes of 120\,pc or more lead to similar, although less dramatic, results because the region size is much larger than the typical diffusion scale length at all energies \citep{DiMauro:2019a}. As discussed in Sect. \ref{nearby}, however, there are several arguments to disfavor such very large extents.

While the observed intensity distributions in the HAWC range can easily be accommodated with confinement region sizes down to 20-30\,pc only, as will be made clear in the next subsections, the local positron flux constraint seems to require much larger sizes, at least for Geminga and in our baseline scenario. This may not be the only solution to the problem, however, and we explore in the following several alternative possibilities to account for the gamma-ray halo properties while being in agreement with the locally measured positron flux, minimizing the extent and magnitude of suppressed diffusion, and getting acceptable injection efficiencies. Before that, we briefly discuss those features in the predicted spectrum of the local positron flux that can guide the search for an acceptable solution. 

\begin{figure}[!t]
\begin{center}
\includegraphics[width=0.9\columnwidth]{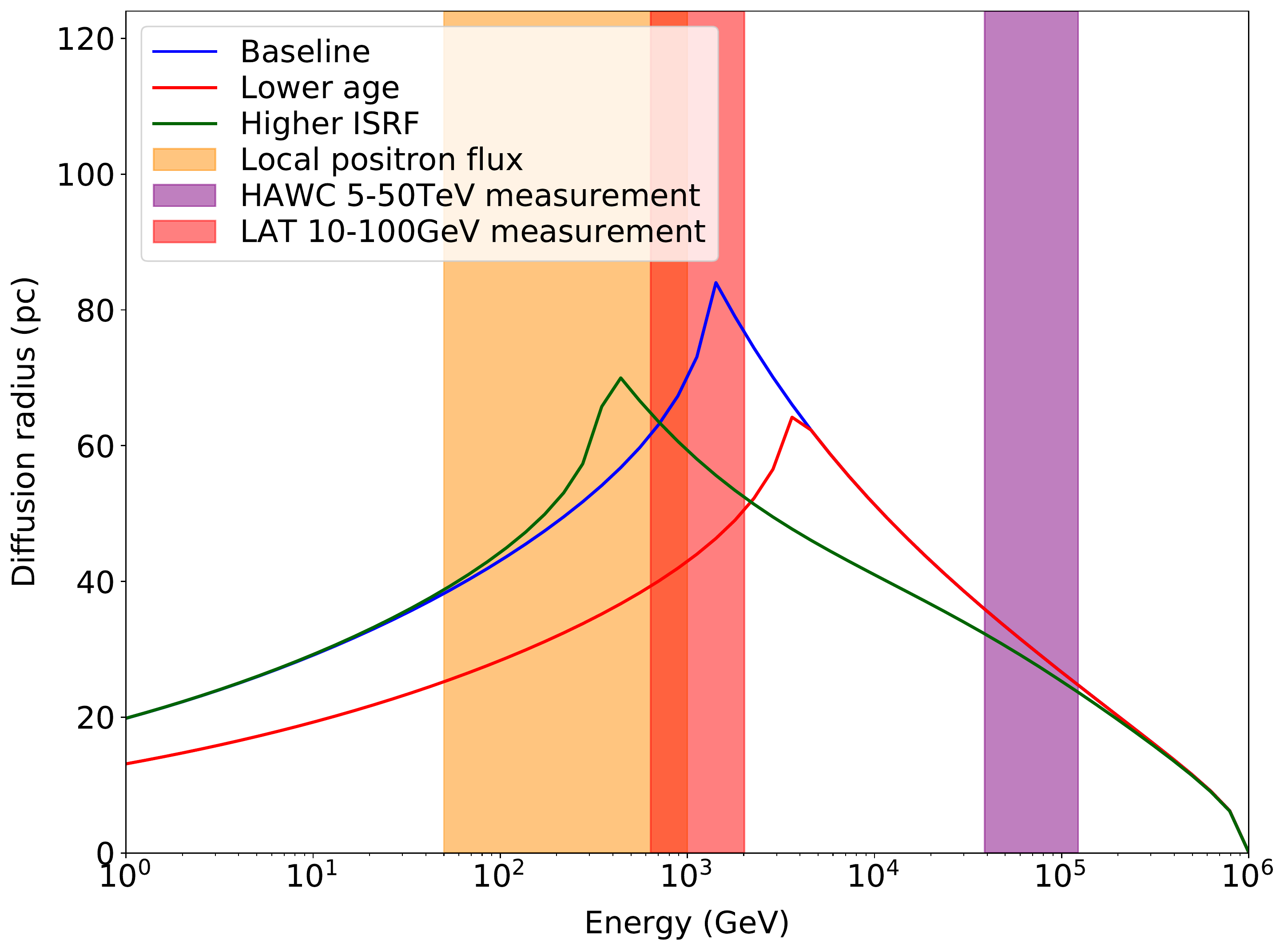}
\caption{Diffusion scale length in three different model setups for J0633+1746: the baseline setup (in blue), an alternative setup with a lower pulsar age (150 instead of 342\,kyr, in red), and a setup with a higher radiation field (8 times stronger than baseline, in green). Overlaid are the typical particle energy ranges probed directly or indirectly by the different observables considered in this work. The AMS-02 experiment is sensitive to positrons with energies from below 1\gev to above 1\tev, but only the range above 50\gev is relevant to this work (at lower energies the flux is dominated by cosmic-ray secondaries).}
\label{fig:mod:diff-radius}
\end{center}
\end{figure}

The shape of the injection spectrum, with a break at 100\gev and harder and softer slopes below and above it, can be recognized in the positron flux spectrum. The low-energy slope is hardly constrained by any of the observables considered here, at least for values in the expected 1.0-2.0 range, while the high-energy slope is constrained to values $\sim2.3-2.5$ in the case of Geminga by the need to reproduce the observed fluxes both in the LAT and HAWC ranges (the non-detection of B0656+14 in the LAT range allows flatter spectra). The peak positron flux reached at about $\sim1-2$\tev, in the case of Geminga, corresponds to the particles with the largest propagation range, resulting from a trade-off between diffusion coefficient and energy losses increasing with energy. Above and below that peak, particles are in loss-limited and diffusion-limited regimes, respectively. This is illustrated in Fig. \ref{fig:mod:diff-radius}, which shows the diffusion radius or diffusion length scale as a function of final energy, for a homogeneous one-zone diffusion scheme. The propagation range of particles in the lower branch is determined by the time allowed for diffusion (from injection start time to pulsar age), while the propagation range of particles in the upper branch is set by the energy loss time (from radiative processes in the magnetic and radiation fields). The normalization of the whole profile is set by the normalization of the suppressed diffusion coefficient, and the momentum dependence of the latter would skew the respective layout of the lower and upper branches. 

The case of two-zone diffusion adds some effects depending on whether the suppressed diffusion region size, which acts as some free escape boundary, lies above or below the diffusion radius. When it is above the diffusion radius (e.g $R_{\textrm{SDR}} = 100-120$\,pc), the particle density at the boundary of the suppressed diffusion region is small and the flux escaping out of it remains modest; conversely, with a suppressed diffusion region smaller than the diffusion radius at some of the energies of interest to us (e.g $R_{\textrm{SDR}} = 20-30$\,pc), the flux leaking out of the halo is higher.

The above discussion points to several possible solutions to reduce the escaping positron flux from the halo of J0633+1746, e.g. enhancing confinement, reducing the time for diffusion, increasing energy losses, or injecting positrons with a flatter spectrum. We examine these possibilities and others in the following subsections.

We note that an additional option would be to assume a much lower diffusion coefficient for the propagation outside the halos, but we chose to not explore that option and consider instead that the coefficient for diffusion in the average \gls{ism} applies locally and is well determined within a factor 2-3, which is much smaller than what would be needed to strongly suppress the flux of positrons from small-sized halos \citep[see][]{Tang:2019}. We also tested an alternative prescription for the momentum dependence of the diffusion coefficient, such as that investigated in \citet{Genolini:2019} to account for spectral features in cosmic-ray direct measurements and where the diffusion coefficient has a steeper momentum dependence below about $250$\,GV.  We implemented the so-called BIG model from that study and found a negligible impact on the predicted signatures.

% Effect of changes in diffusion suppression
\subsection{Effect of changes in diffusion suppression}
\label{res:effdiff}

\begin{figure}[!t]
\begin{center}
\includegraphics[width=0.9\columnwidth]{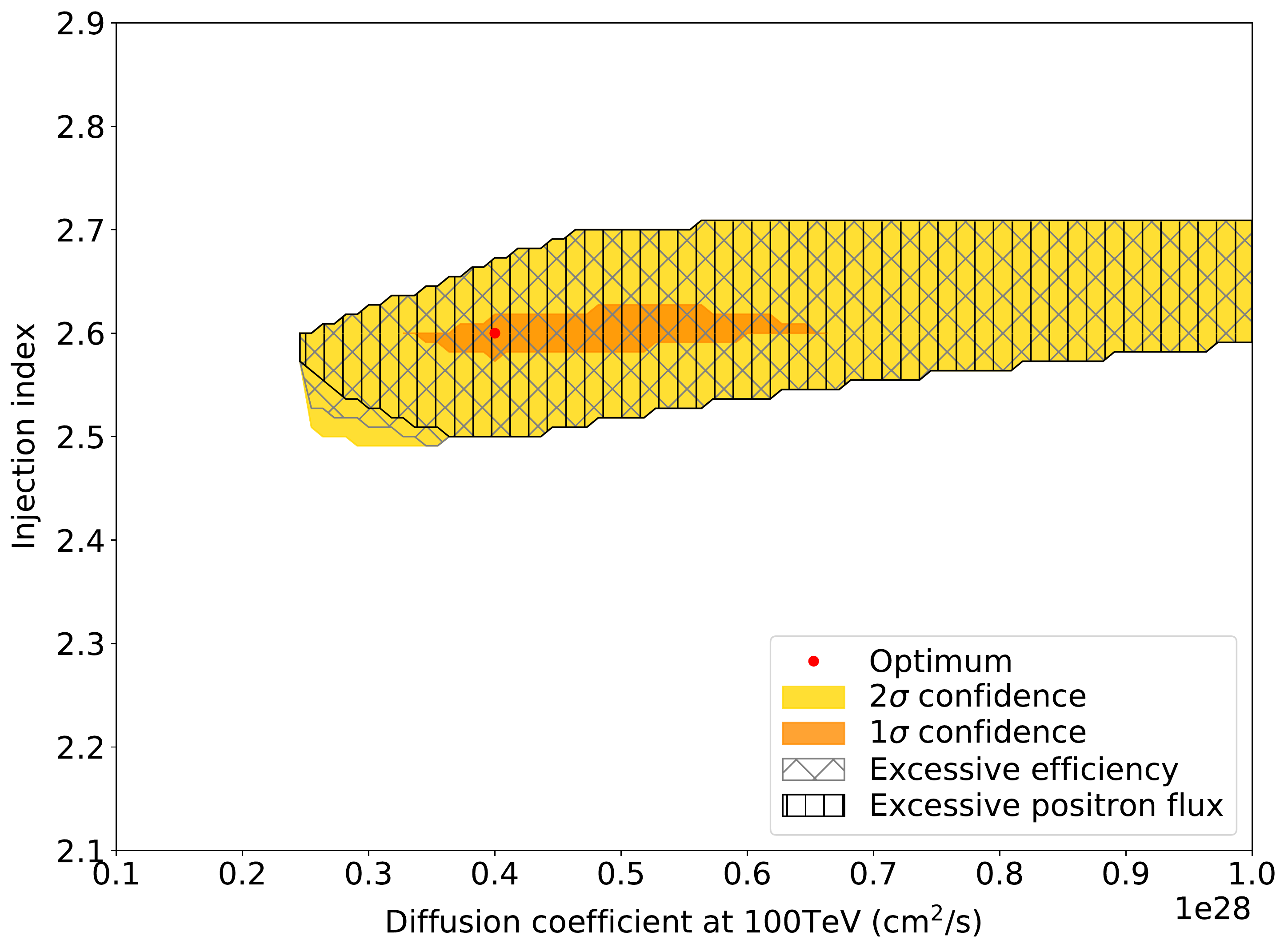}
\includegraphics[width=0.9\columnwidth]{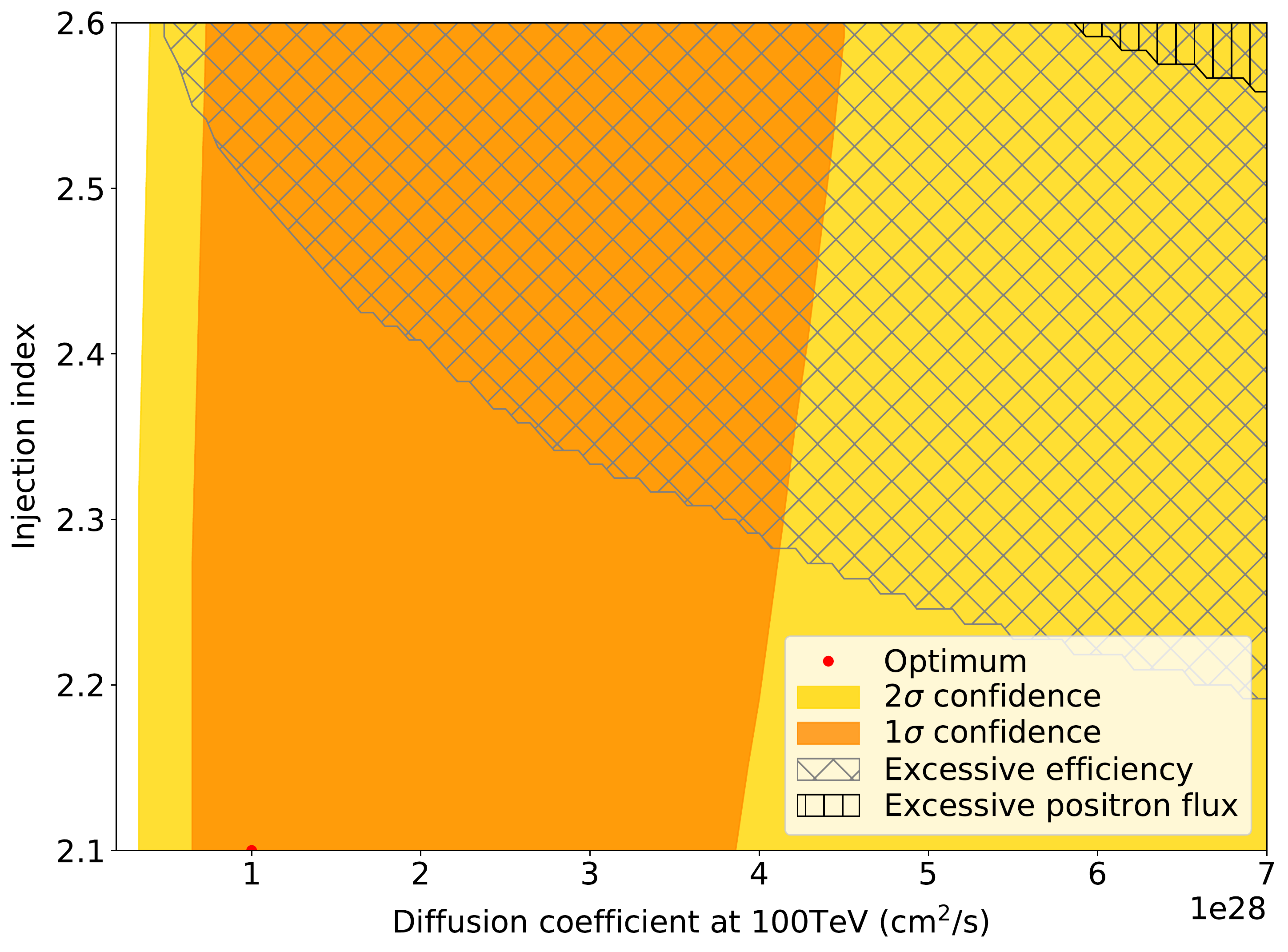}
\caption{Confidence interval in the diffusion normalisation and injection index space, for model fits to HAWC+LAT data in the case of J0633+1746, and to HAWC data only in the case of B0656+14. The model setups feature a suppressed diffusion region extent of 50\,pc. The color coding indicates increasing $\chi^2$ variations from the optimum, by up to 1 and 4 in the orange and yellow regions, respectively. Black vertical hatches mark the model fits yielding positron fluxes in excess of the AMS-02 measurement in the 0.1-1\tev range, while gray slanting hatches mark model fits implying injection efficiencies in excess of 100\% of the spin-down power.}
\label{fig:mod:confidence:diff-idx}
\end{center}
\end{figure}

We first examine the uncertainty range on the diffusion coefficient to determine how much the magnitude of diffusion suppression can be relaxed in our baseline model. We first note that a solution to suppress the positron flux at $\sim0.1-1$\tev while still fitting the intensity profile in the HAWC energy range can be achieved by increasing the momentum dependence of the suppressed diffusion coefficient. Although not reported here, we explored such a possibility and concluded that such a choice results in a non-negligible contribution to the positron flux and actually shifts the problem to that of explaining very high suppression factors $\gtrsim2000$ at $\lesssim1$\tev energies\footnote{A harder momentum dependence of the diffusion coefficient, with a power-law index increasing from 0.3 to 0.8, can actually bring the positron flux below the $\sim1$\tev observed level for values of 0.6-0.7. The effect is non-linear and depends on the relative values of the energy-dependent diffusion radius and suppressed diffusion region size at the relevant energy. When the former approaches and drops below the latter, the particle density at the halo boundary is progressively cut off exponentially \citep[the {\it erfc} term prevails over the {\it 1/r} dependence in  Eq. S4 in][]{Abeysekara:2017b}, and so do the resulting escaping fluxes. In the baseline scenario, the 1\tev diffusion radius is of order 80\,pc and bringing it down to a 50\,pc size for the confinement region can be achieved by reducing the diffusion coefficient at 1\tev by about 3 (the diffusion radius depends on the square root of the diffusion coefficient). Such a reduction can be obtained using a diffusion index of 0.6 instead of 0.3. If the confinement region is 30\,pc in size instead, matching the 1\tev diffusion radius requires reducing the diffusion coefficient by about 7, which in turn implies a diffusion index of 0.8 instead of 0.3. Such stronger momentum dependence are not unexpected theoretically in the self-confinement scenario \citep[see Fig. 2 in][]{Evoli:2018}, at least over a selected range of distances from the pulsar and times since particle/turbulence injection.}.

The diffusion suppression level actually is pretty well constrained for both objects, as illustrated in Fig. \ref{fig:mod:confidence:diff-idx}. For models with suppressed diffusion region extents of 50\,pc, and from the HAWC and LAT constraints alone, the diffusion coefficients could be increased up to nearly $10^{28}$ and $10^{29}$\dunit\ at 100\tev, for J0633+1746 and B0656+14, respectively. Yet, this has to be compensated by higher and higher injection efficiencies, rapidly reaching 100\% for B0656+14 and largely exceeding 200\% for J0633+1746, making the violation of the positron flux constraint even more severe in the latter case. On the other side, smaller values of the diffusion coefficient do not help as they rapidly lead to degraded fits to the HAWC intensity profile.

Eventually, the most straightforward solution to the excessive positron flux issue in the baseline model setup is to enlarge the suppressed diffusion region, as has been explored in other works already \citep[e.g.][]{DiMauro:2019a}. More extended suppressed diffusion regions place the boundary at positions where the particle density is lower, especially when it is beyond the diffusion length scale, which reduces the escaping flux as illustrated in Figs. \ref{fig:mod:geminga:effect-size} and \ref{fig:mod:monogem:effect-size}. This is going in the opposite direction of what we are trying to achieve, i.e. minimizing the conditions of diffusion suppression, but it remains a physically viable solution.

In the case of Geminga, satisfactory fits to the observables required reducing the high-energy injection index, which brought the injection efficiency down to acceptable values. Sizes of 80\,pc and above provide good fits to all observables, although the 80\,pc case nearly saturates the 1\tev positron flux inferred from the latest AMS-02 measurements \citep{Aguilar:2019a}. In the case of B0656+14, the baseline confinement region size of 50\,pc already provided a good fit to all observables, and so does the 30\,pc case, as illustrated in the bottom panel of Fig. \ref{fig:mod:monogem:effect-size}.

% Effect of changes in the magnetic/radiation fields
\subsection{Effect of changes in the magnetic/radiation fields}
\label{res:efffields}

\begin{figure}[!t]
\begin{center}
\includegraphics[width=0.9\columnwidth]{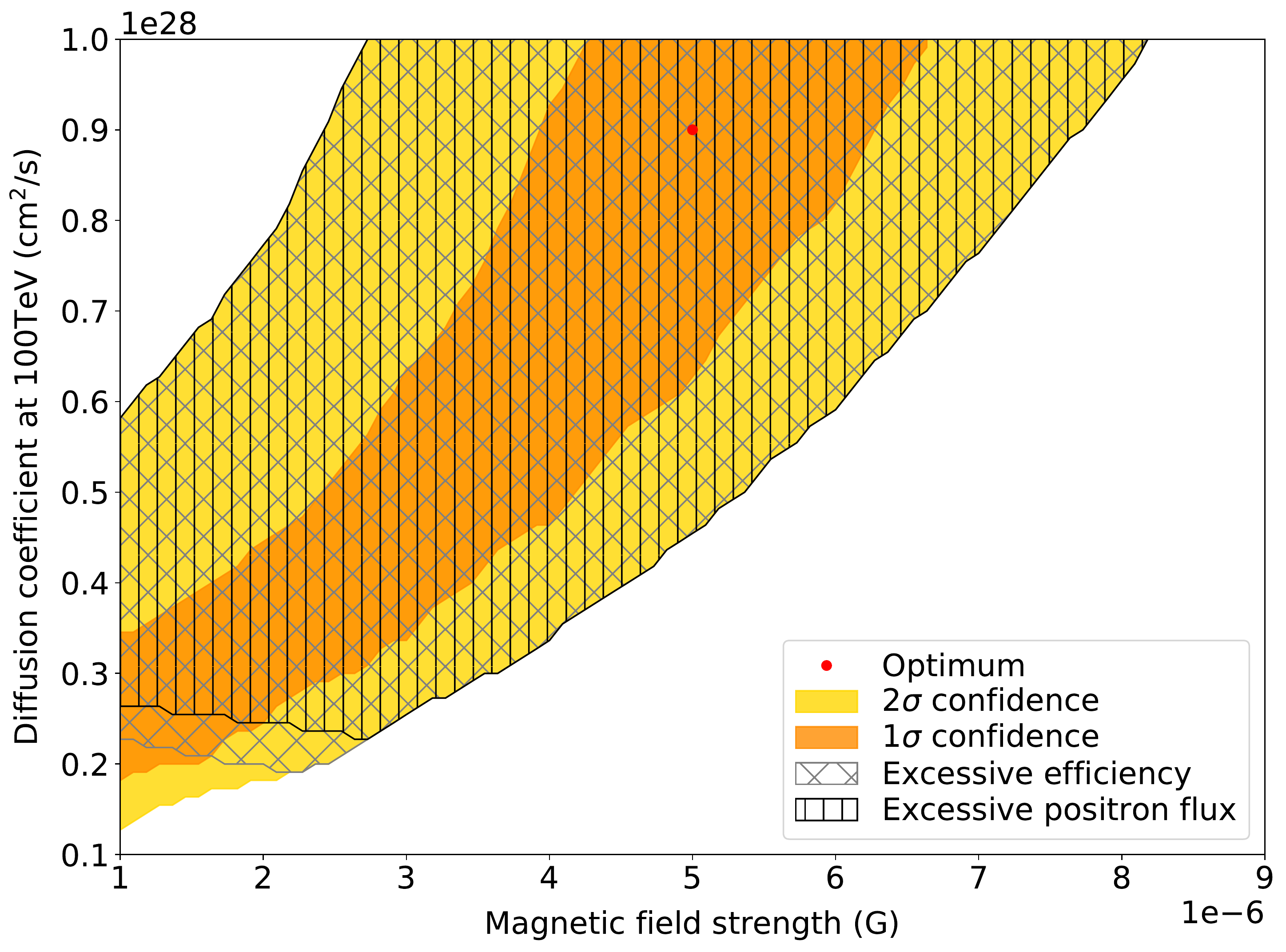}
\includegraphics[width=0.9\columnwidth]{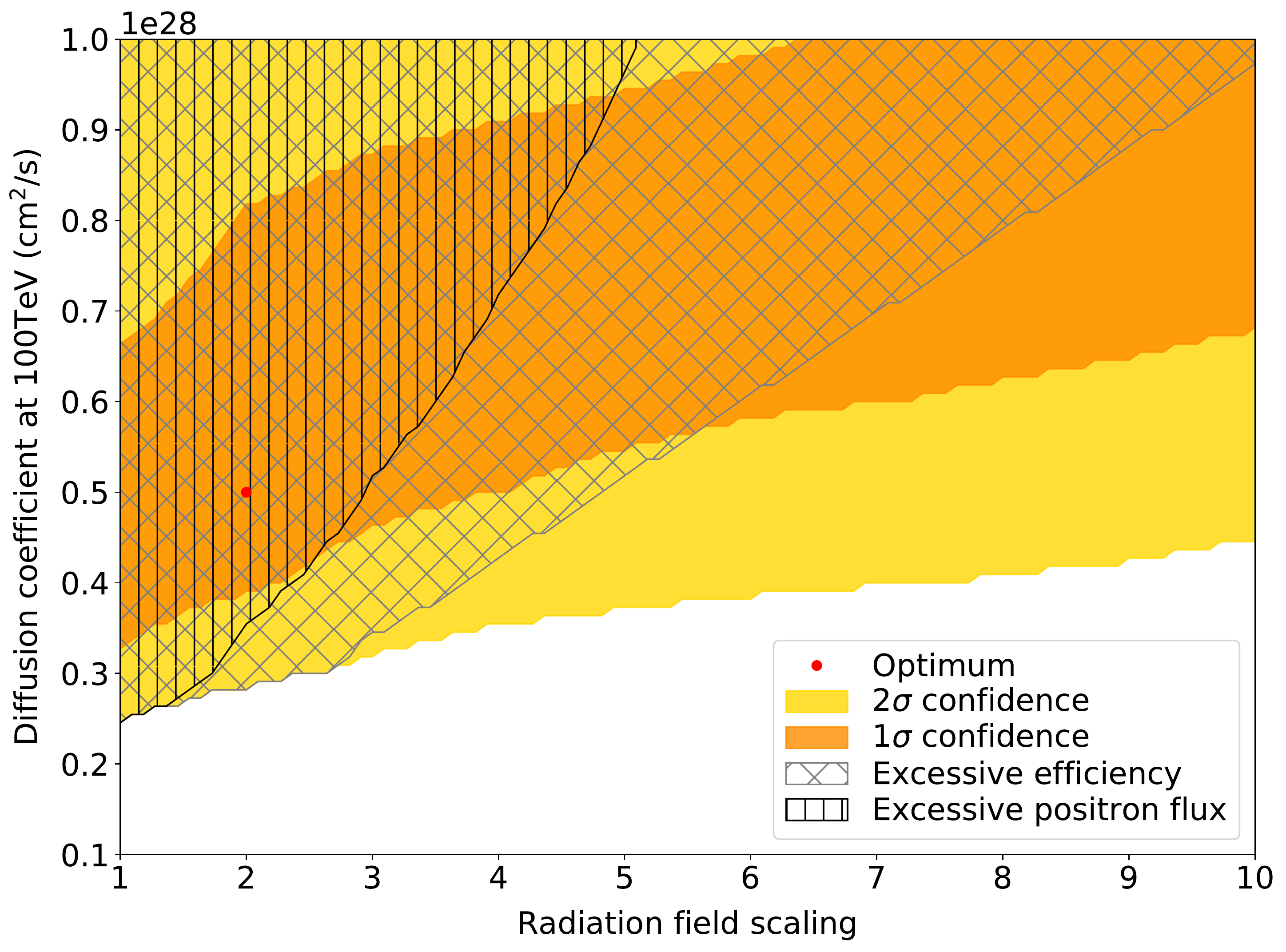}
\caption{Confidence interval in the diffusion normalisation and magnetic field strength space (top panel), and diffusion normalisation and radiation field intensity space (bottom panel), for model fits to HAWC+LAT data in the case of J0633+1746, assuming a suppressed diffusion region extent of 50\,pc. The meaning of graphical elements is the same as in Fig. \ref{fig:mod:confidence:diff-idx}.}
\label{fig:mod:confidence:diff-fields}
\end{center}
\end{figure}

As illustrated in Fig. \ref{fig:mod:confidence:diff-idx}, higher diffusion coefficients lead to progressively degraded fits to the HAWC data because the corresponding intensity distributions are more spread out. This trend can be compensated with: (i) a larger distance to the pulsar, to fit a more diluted physical distribution into the observed angular distribution; (ii) higher energy losses, to reduce the larger diffusion scale length. We examine here the latter option, while the former one will be dealt with in the next subsection.

Compensating for a higher diffusion coefficient with higher energy losses from a stronger magnetic field indeed allows us to fit the HAWC profile, as illustrated in the top panel of Fig. \ref{fig:mod:confidence:diff-fields}. In the case of J0633+1746, variations in the magnetic field of up to $\sim10$\,$\mu$G allows for the diffusion coefficients to be relaxed up to $10^{28}$\dunit\ or more. Yet, this comes with ever increasing injection efficiencies, to ensure that sufficient energy is channeled into inverse-Compton scattering, which eventually translates into a larger and still excessive positron flux.

\begin{figure}[!t]
\begin{center}
\includegraphics[width=0.9\columnwidth]{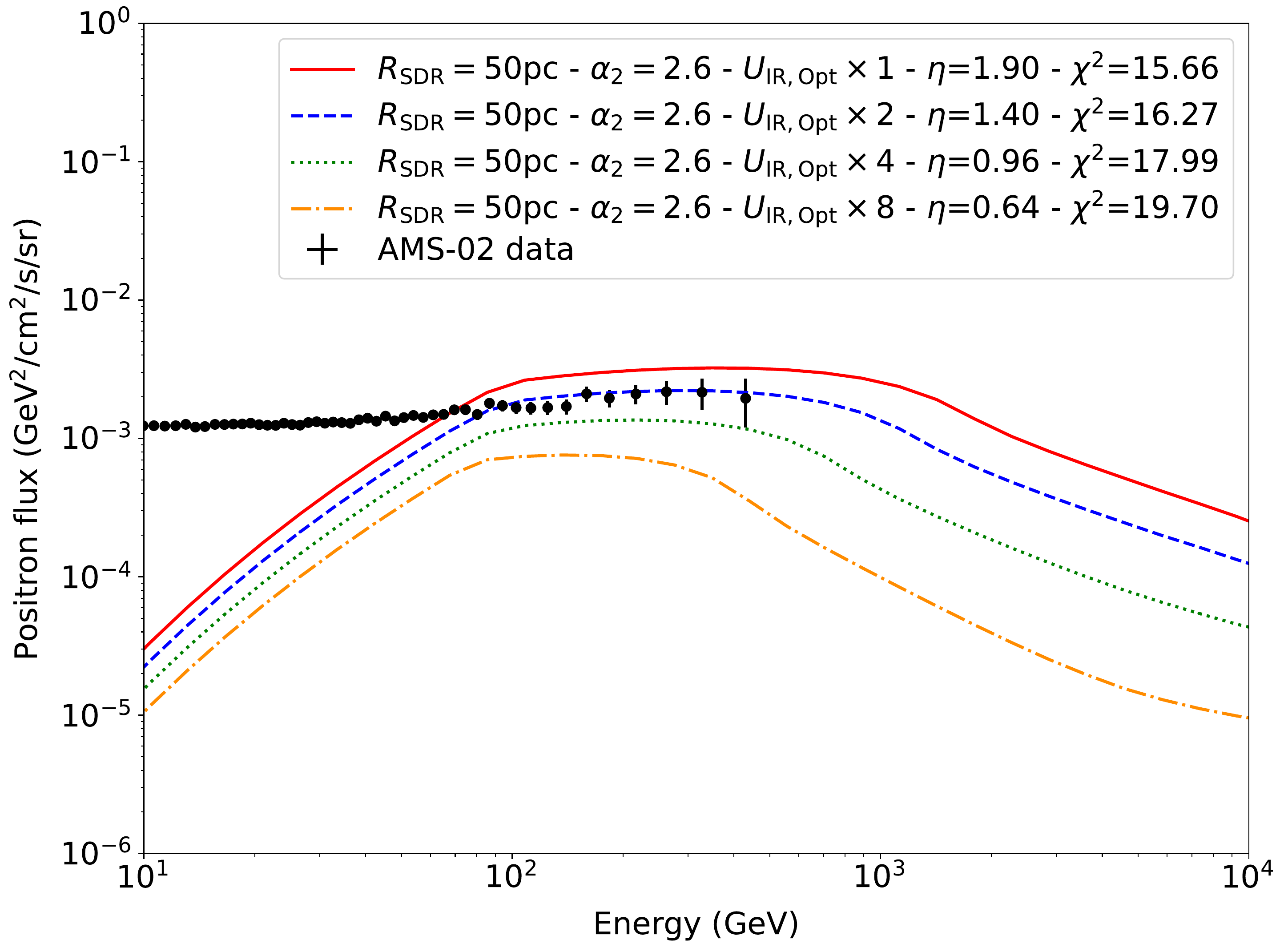}
\includegraphics[width=0.9\columnwidth]{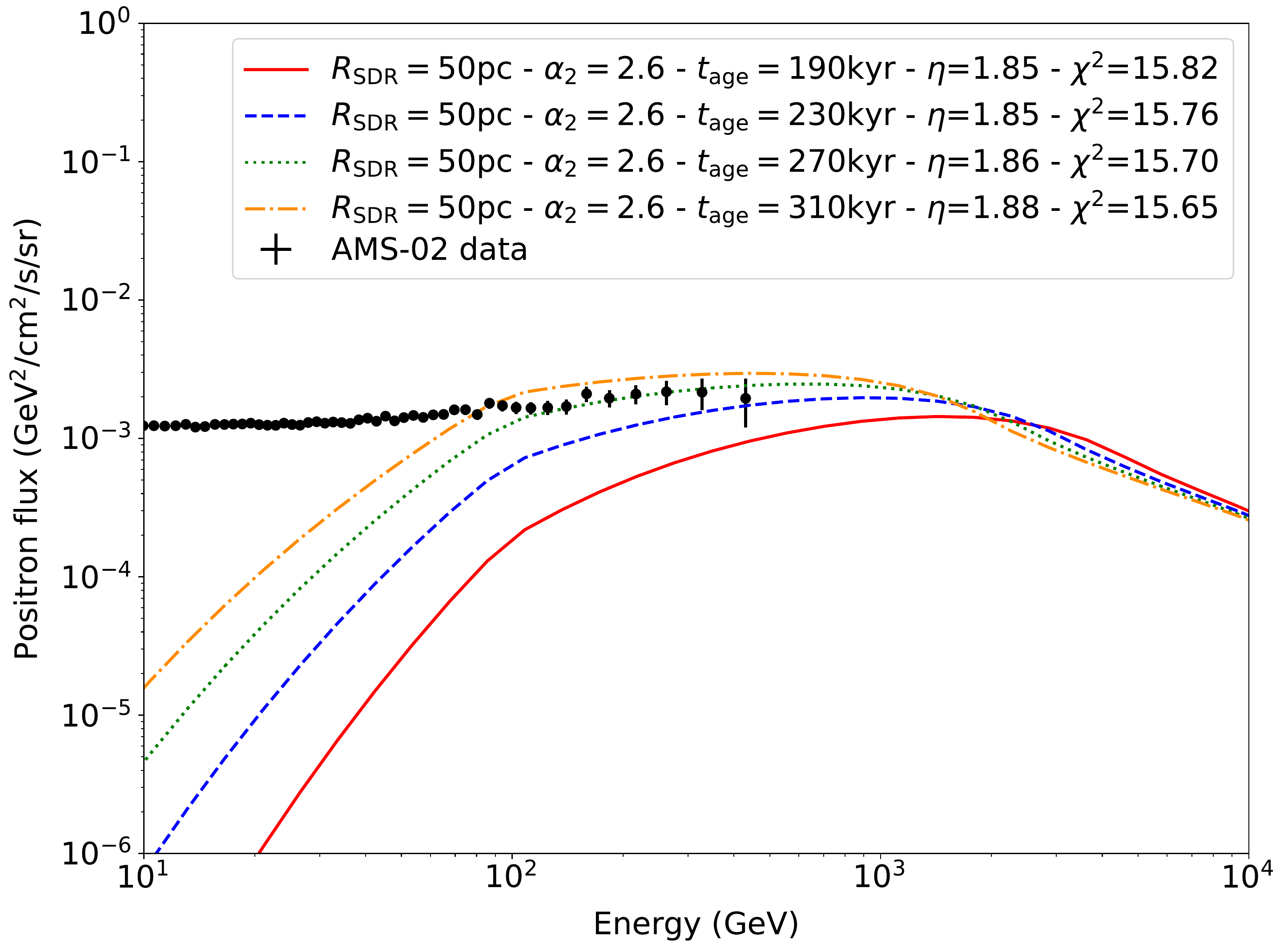}
\caption{Predicted local positron fluxes for different increases of the infrared and optical photon field densities $U_{\rm IR,Opt}$ (top panel) and different ages for the pulsar $t_{\rm age}$ (bottom panel), assuming a suppressed diffusion region extent of $R_{\textrm{SDR}}=50$\,pc. The predicted gamma-ray emission is fitted to HAWC+LAT observations of Geminga and the corresponding injection efficiencies $\eta$ and $\chi^2$ of the fits are given in the legend.}
\label{fig:mod:effect-radfield-psrage}
\end{center}
\end{figure}

Increasing instead the intensity of the radiation fields involved in inverse-Compton scattering (except the cosmic microwave background) produces viable solutions for two reasons: (i) when increasing the field intensities, a given level of gamma-ray emission in the HAWC or LAT band can be reproduced with a smaller injection efficiency, hence a correspondingly smaller flux of diffusion-limited particles and even smaller flux of loss-limited particles; (ii) enhanced energy losses produce a shift of the positron flux peak toward lower energies and smaller fluxes. The net effect is illustrated in Figs. \ref{fig:mod:diff-radius} and \ref{fig:mod:effect-radfield-psrage}, while the allowed parameter space is shown in the bottom panel of Fig. \ref{fig:mod:confidence:diff-fields}.

Radiation fields a few times stronger than our baseline assumption can solve the issues of excessive local positron flux and injection efficiency in the model setup with a 50\,pc suppressed diffusion region applied to Geminga. Substantiating this possibility would require an investigation of the stellar content in the vicinity of the pulsar, to establish whether it is conceivable that the radiation field be at least 5-6 times higher than the average for the Galaxy at this position \citep[from e.g. the model of][]{Popescu:2017}. Such enhancements however leave a significant contribution of Geminga to the $\sim0.1-1$\tev positron flux and only allow for a modest relaxing of the diffusion coefficient as it needs to be compensated by a higher injection efficiency, which raises the positron flux above the AMS-02 measurement again. 

% Effect of changes in pulsar parameters
\subsection{Effect of changes in pulsar parameters}
\label{res:effpsr}

\begin{figure}[!t]
\begin{center}
\includegraphics[width=0.9\columnwidth]{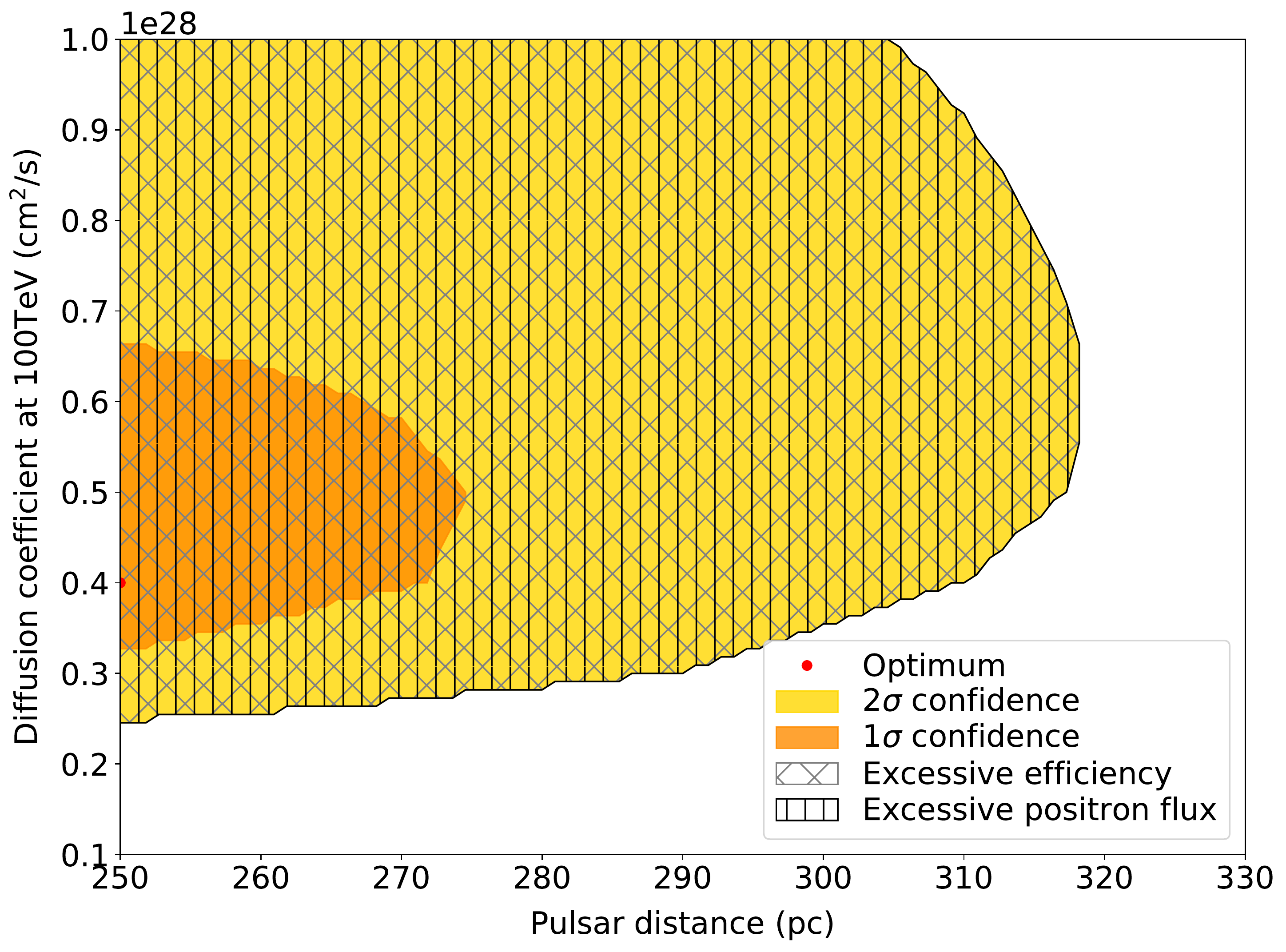}
\includegraphics[width=0.9\columnwidth]{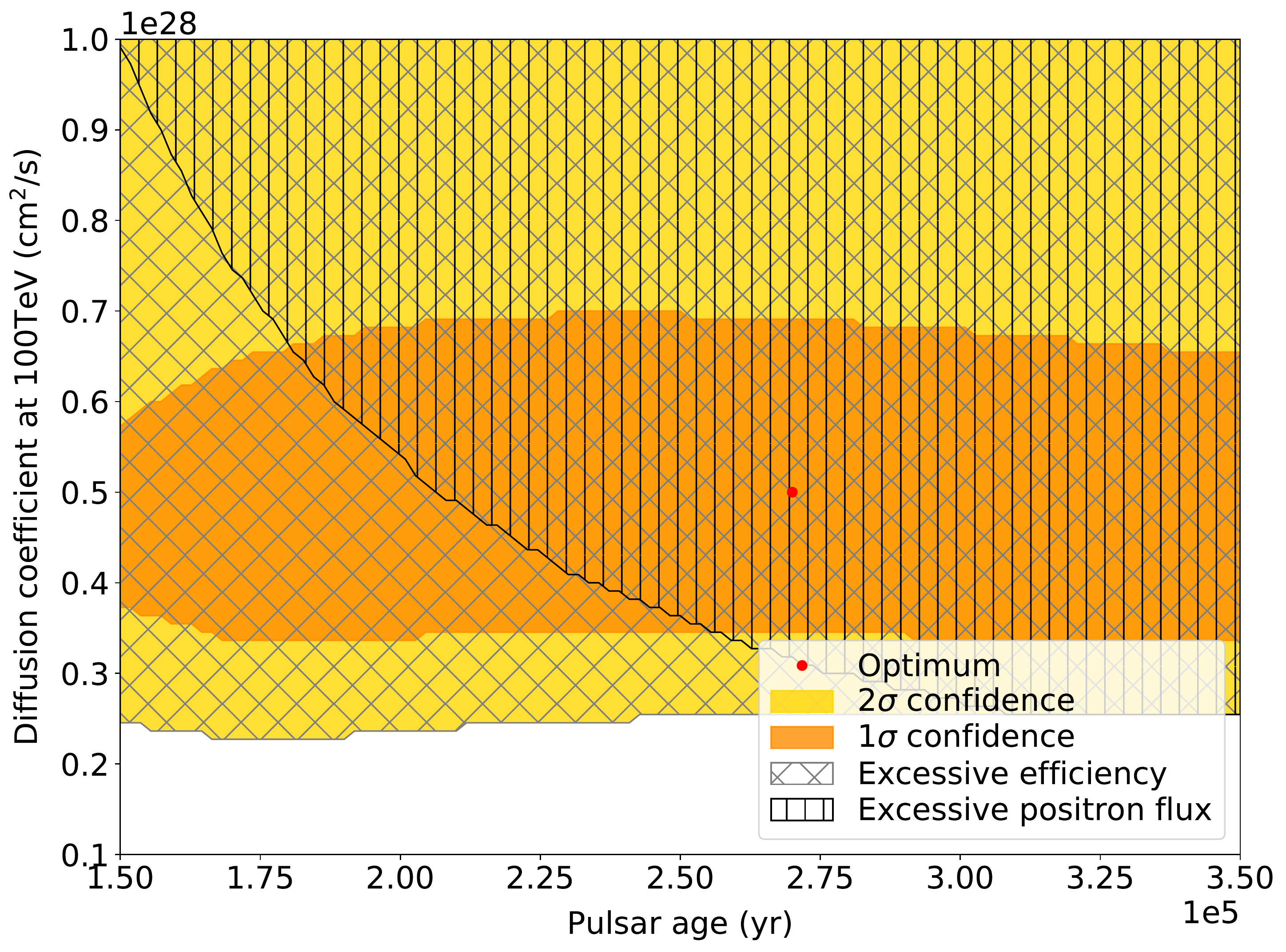}
\caption{Confidence interval in the diffusion normalisation and pulsar distance space (top panel) and diffusion normalisation and pulsar age space (bottom panel), for model fits to HAWC+LAT data in the case of J0633+1746, assuming a suppressed diffusion region extent of 50\,pc and injection index of 2.6. The meaning of graphical elements is the same as in Fig. \ref{fig:mod:confidence:diff-idx}.}
\label{fig:mod:confidence:diff-psr}
\end{center}
\end{figure}

Higher diffusion coefficients lead to intensity distributions that are more spread out, an effect that needs to be compensated in some way to fit a more diluted physical distribution into the observed angular distribution. A larger distance to the pulsar is a possible solution.

In the case of Geminga, the upper bound of the distance range published in \citet{Verbiest:2012} is 480\,pc, nearly twice the baseline 250\,pc used here. Yet, as illustrated in the top panel of Fig. \ref{fig:mod:confidence:diff-psr}, a larger distance only allows for a modest increase in the diffusion coefficient when the suppressed diffusion region has a 50\,pc size. This comes from the fact that the particles radiating in the HAWC range have a diffusion scale length already close to 50\,pc (see Fig. \ref{fig:mod:diff-radius}), such that relaxing the diffusion coefficient from its baseline value does not significantly modify the intensity distribution within 50\,pc of the pulsar. The effect is more significant when the suppressed diffusion region has a 80\,pc size. Yet, in both cases, the possible solutions lead to a larger positron flux because the higher diffusion coefficient, and to a lesser extent the larger distance, need to be compensated by a higher injection efficiency still largely in excess of 100\%.

Another solution to suppress the excessive predicted positron flux below 1\tev is to reduce the time allowed for diffusion. For lower-energy particles in the diffusion-limited regime, this implies reducing the time since injection start. This could be done with a delayed particle release from the pulsar, but we already use a rather high 60\,kyr baseline value. Instead, a significant effect can be obtained by assuming a smaller age for the Geminga pulsar. 

The 342\,kyr value used as baseline actually corresponds to the characteristic age and is based on the assumption of: (i) spin-down driven by magnetic dipole radiation, with a braking index of 3; (ii) the unknown initial spin period being much smaller than the present-day value. Both assumptions may well be inappropriate for the Geminga pulsar, as they were shown to be for other pulsars. In \citet{Suzuki:2021}, the characteristic ages of pulsars were shown to overestimate alternative, presumably more reliable, historical, light-echo, or kinematic age estimates of their parent systems, by factors of a few or more. 

The bottom panels of Figs. \ref{fig:mod:effect-radfield-psrage} and \ref{fig:mod:confidence:diff-psr} show that ages $\lesssim 250$\,kyr result in a predicted positron flux in agreement with AMS-02 data, while still allowing for solutions consistent with HAWC and {\it Fermi}-LAT observations. This, however, does not solve the excessive injection efficiency, which remains close to 200\%. Eventually, the best solution to that issue, as investigated in the next section, is to reduce the injection index, at the expense of a poorer fit to the {\it Fermi}-LAT data. For a true pulsar age of 190\,kyr, reducing the injection index from 2.6 to 2.5 allows for a decent fit to all observables with an injection efficiency of 110\%.

A smaller age for the Geminga pulsar, in the $150-200$\,kyr range, has interesting consequences on the birth place of the pulsar, which would be closer to the Galactic mid-plane, and on the morphology of the halo at low energies, with trailing emission at low GeV energies that should be less extended. The latter point would have to be tested against {\it Fermi}-LAT data, although the very large extension of the detected source and its limited significance may preclude disentangling halo models with different pulsar ages \citep{DiMauro:2019a}.

% Effect of changes in injection spectrum
\subsection{Effect of changes in injection spectrum}
\label{res:effinj}

\begin{figure}[!t]
\begin{center}
\includegraphics[width=0.9\columnwidth]{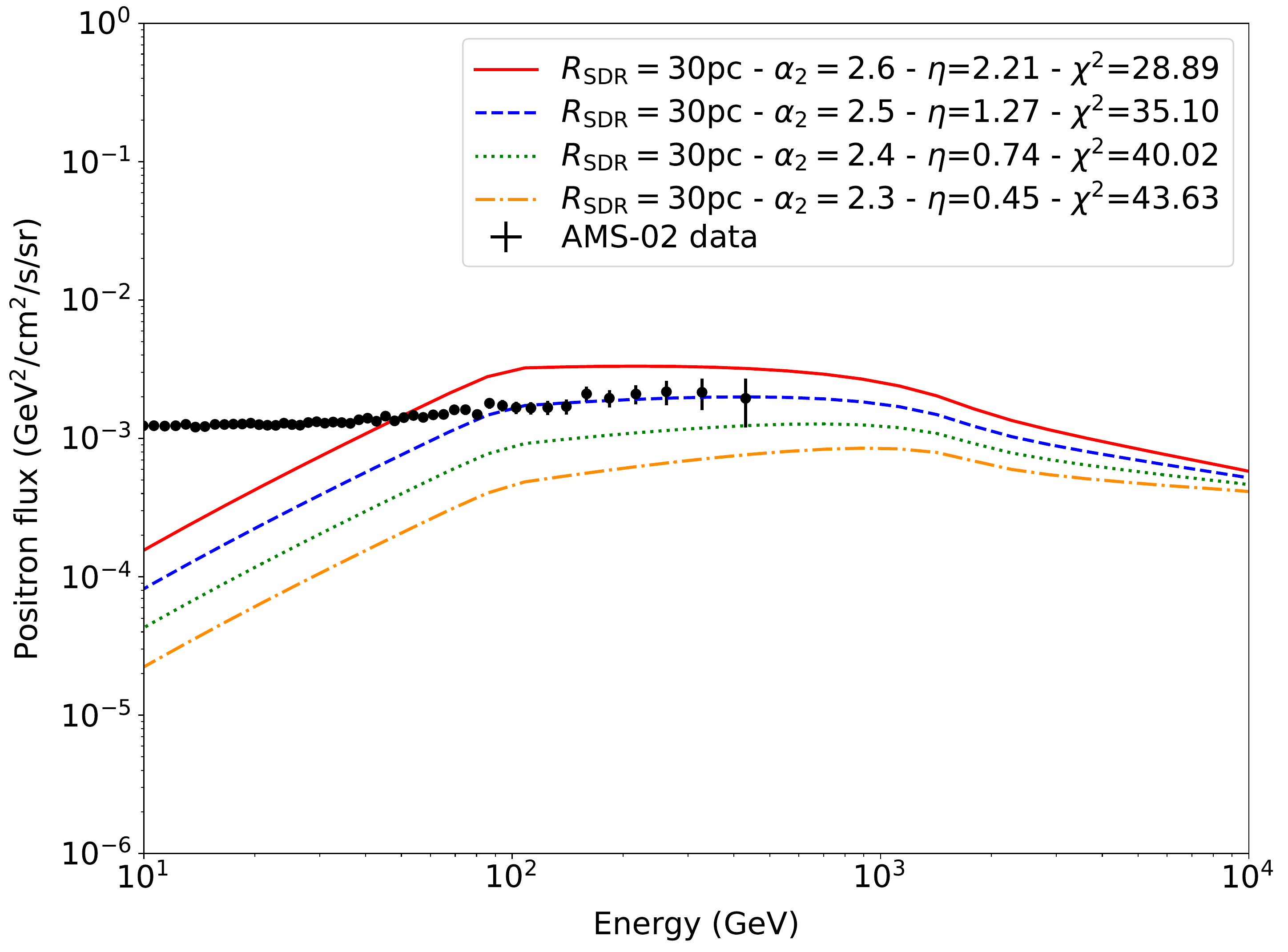}
\caption{Predicted local positron fluxes for variations of the injection index $\alpha_2$, assuming a suppressed diffusion region extent $R_{\textrm{SDR}}=30$\,pc. The predicted gamma-ray emission is fitted to HAWC observations of Geminga only and the corresponding injection efficiencies $\eta$ and $\chi^2$ of the fits are given in the legend. The $\chi^2$ includes the deviation from the LAT spectral points to illustrate the effect of neglecting that constraint.}
\label{fig:mod:effect-varinj}
\end{center}
\end{figure}

In the case of Geminga, reproducing the HAWC+LAT constraints for a confinement region of 50\,pc, without exceeding the AMS-02 measurements and while getting an injection efficiency below 100\%, turned out to be quite challenging, even with all the options explored above. Beyond a smaller pulsar age or higher radiation field density, an alternative solution can be obtained by relaxing one of the constraints. In particular, motivated by the result of \citet{Xi:2019a}, we considered the possibility that the {\it Fermi}-LAT measurement in \citet{DiMauro:2019a} is overestimating the GeV flux, and turned it into an upper limit instead. The situation then becomes similar to that of B0656+14, for which only upper limits were obtained from LAT observations.

Acceptable solutions for Geminga are then easily found with injection indices of 2.4 or below, and suppressed diffusion regions as small as 30\,pc, nearly the minimum value allowed by the HAWC intensity profile. This solves the excessive positron flux issue, as illustrated in Fig. \ref{fig:mod:effect-varinj}, and leads to injection efficiencies of about 70\% or below. Interestingly, such injection efficiencies are comparable to those needed for B0656+14 to match the HAWC constraint with a similar model setup featuring suppressed diffusion regions of $30-50$\,pc (see Fig. \ref{fig:mod:monogem:effect-size}).

Relaxing the LAT constraint actually may allow for an even more drastic modification of the injection spectrum, by making it possible to drop the broken power-law assumption in favour of a single power-law. In such a scenario, the relatively soft emission in the HAWC band can be reproduced by shifting the cutoff from our default 1\pev energy to smaller values in the $10-100$\tev range. Such a low-energy cutoff has the effect of softening the spectrum of particles radiating in the $\sim5-50$\tev HAWC band, while allowing for a hard injection spectrum overall, with indices in the $1.5-2.0$ range. This strongly suppresses the contribution to the local positron flux and requires a very small injection efficiency. The same logic applies to B0656+14 and is reminiscent of the work presented in \citet{Hooper:2017}. The extension of gamma-ray measurements to 100\tev energies and above will tell whether this is a viable option \citep{Sudoh:2021}.

% Summary
\subsection{Summary}
\label{res:summary}

We summarize here the different model setups found to yield acceptable fits to all (or a subset of all) observables for J0633+1746 and B0656+14, under the constraint of minimizing the extent and magnitude of diffusion suppression as much as possible. 

In the case of J0633+1746 and suppressed diffusion regions of moderate sizes $\lesssim80$\,pc, the highly suppressed diffusion coefficient is constrained by both the HAWC and AMS-02 observations and the total energetics of the pulsar. Eventually, the following model setups were found to provide acceptable fits: (i) a 80\,pc suppressed diffusion region size with injection index $2.3-2.4$; (ii) a 50\,pc suppressed diffusion region size with injection index $2.5-2.6$, assuming a true pulsar age $\lesssim200$\,kyr; (iii) a 30\,pc suppressed diffusion region size with injection index $2.3-2.4$, relaxing the constraint to fit the claimed detected flux in the {\it Fermi}-LAT range. These scenarios feature a suppressed diffusion coefficient in the range $2-6 \times 10^{27}$\dunit\ at 100\tev, and involve injection efficiencies in the range $50-110$\%.

The case of B0656+14 is less constrained, owing to the non-detection in the {\it Fermi}-LAT range and smaller age and larger distance of the pulsar. These conspire to yield a smaller contribution to the local positron flux, which makes it easier to fit within the AMS-02 measurement. Eventually, we retained the following model setups: (i) a 50\,pc suppressed diffusion region size with injection index $2.1-2.5$; (ii) a 30\,pc suppressed diffusion region size with injection index $2.1-2.3$. These scenarios feature a suppressed diffusion coefficient in the range $1-6 \times 10^{28}$\dunit\ at 100\tev, and involve injection efficiencies in the range $60-100$\%.

To limit the computational cost of the work, we did not explore all possible combinations of parameters but it is clear that there exists more model setups than listed above that could provide similar or even better fits to the observations. For instance, in the case of Geminga, a smaller age for the pulsar combined with a 30\,pc suppressed diffusion region size and the appropriate injection index to fit (or not) the {\it Fermi}-LAT spectrum could be an option.

In all the above scenarios, we emphasize that our assumption of late injection, past a \gls{pwn} stage lasting for 60\,kyr, clearly helps to reduce the predicted positron flux. Injection starting earlier leads to a higher output of $\lesssim 1$\tev particles, and a bit more time for diffusion, which makes it even more difficult to match all three observables. The important point is that, in all cases, significant contributions to the $\gtrsim1$\tev local positron flux are expected from J0633+1746 and B0656+14, unless the suppressed diffusion regions are extremely large.

Future high-precision measurements of the positron flux in the few TeV range may offer a way to discriminate between all possible scenarios considered here. Observations in the {\it Fermi}-LAT core band also is an obvious way to tell possibilities apart, but we are probably at the limit of what can be done for a very extended source like Geminga. Potentially more promising are the implications at the Galactic population level of the different confinement region sizes in these scenarios. All other things being equal, especially a typical distribution for injection efficiencies from pulsars/\gls{pwne}, a larger suppressed diffusion region leads to higher gamma-ray fluxes, and to a brighter population if most or all middle-aged pulsars go through a halo phase similar to J0633+1746 or B0656+14. The latter aspect will be explored in a subsequent publication, while the following section will deal with the implications on the positron flux of a putative local population of pulsar halos.

% Positron flux from nearby pulsars
\section{Positron flux from nearby pulsars}
\label{nearby}

In this section, we assess the consequences in terms of local positron flux of a scenario where all nearby pulsars around us develop a halo similar to J0633+1746 or B0656+14. This is complementary to a full Galactic pulsar halo population synthesis, an example of which will be presented in a forthcoming paper.

We extract from the ATNF data base \citep{Manchester:2005} a selection of pulsars with estimated distances $\leq$1\,kpc, characteristic ages $\geq60$\,kyr and spin-down power $\geq 10^{33}$\punit. This amounts to 16 objects including J0633+1746 and B0656+14, which we exclude from the sample as they were treated separately already, taking constraints from their detected gamma-ray emission into account. For the remaining 14 objects, we compute the predicted positron flux for the different halo model setups selected in Sect. \ref{res:summary}.

\begin{figure}[!t]
\begin{center}
\includegraphics[width=0.9\columnwidth]{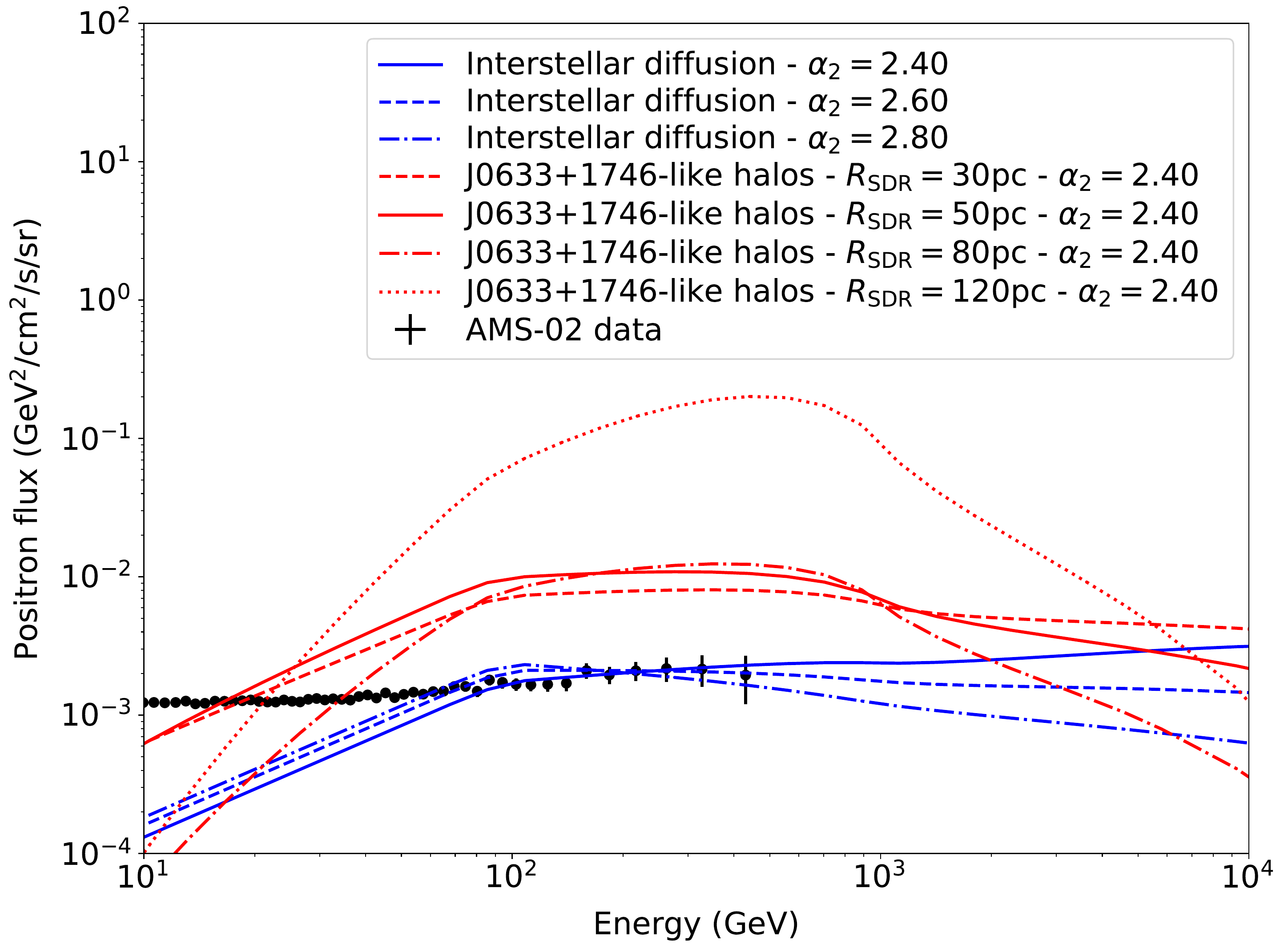}
\includegraphics[width=0.9\columnwidth]{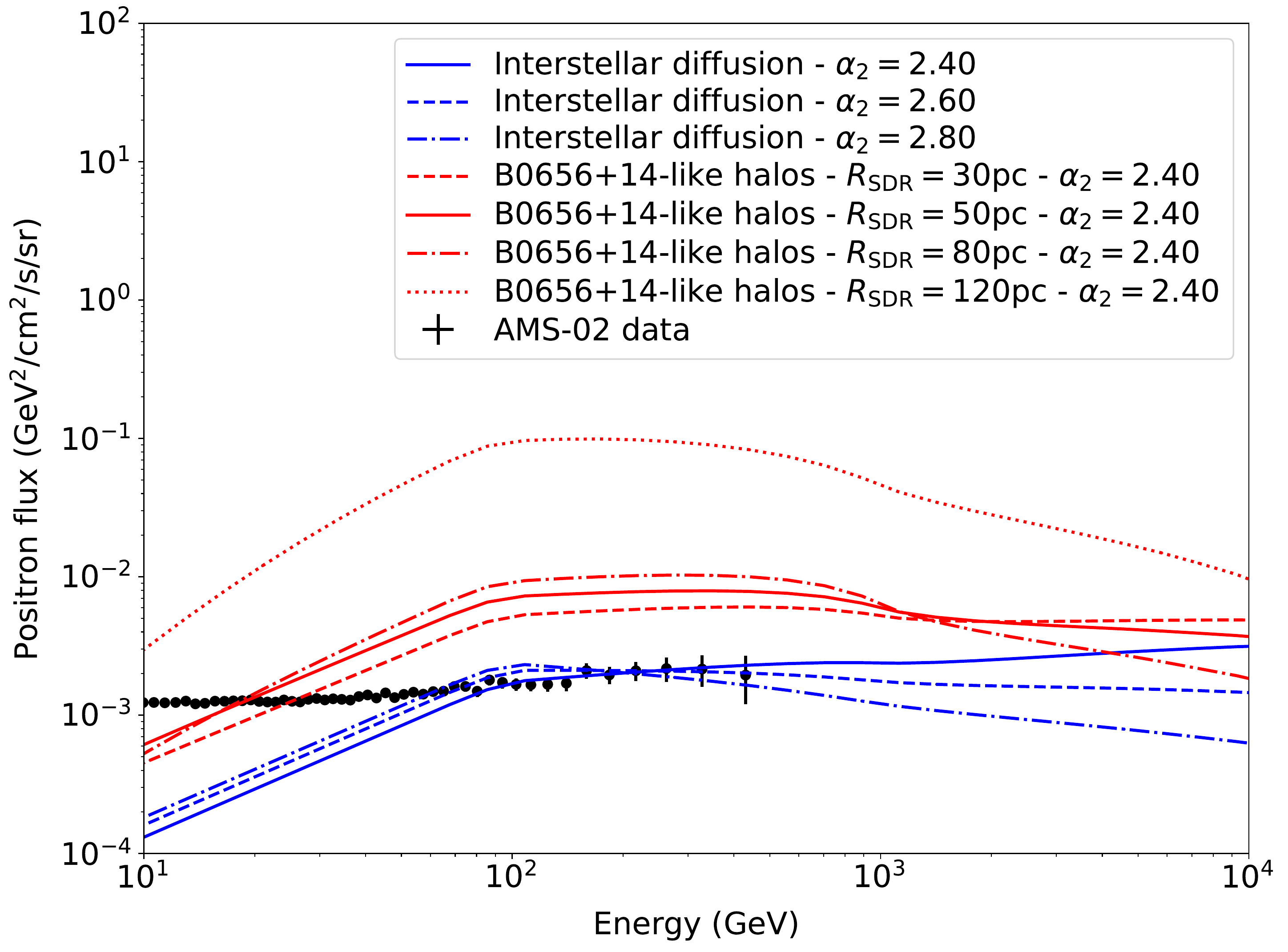}
\caption{Total local positron flux from known middle-aged pulsars within 1\,kpc, except J0633+1746 and B0656+14, under the assumption that they develop a halo like J0633+1746 (top panel) or B0656+14 (bottom panel) past an age of 60\,kyr, with different possible sizes for the suppressed diffusion region. For comparison, both plots also display the expected flux if all nearby pulsars develop no halo and particles released in the \gls{ism} diffuse out without experiencing any kind of strong and extended confinement around the source. By default, injection efficiencies of 100\% are assumed, and several possibilities were tested for the injection index in the case of plain interstellar diffusion.}
\label{fig:nearby:flux}
\end{center}
\end{figure}

Figure \ref{fig:nearby:flux} shows the total local positron flux from all other known nearby middle-aged pulsars, assuming they develop a halo similar to J0633+1746 or B0656+14. As typical properties for the latter, we assumed a suppressed diffusion coefficient of $4 \times 10^{27}$\dunit\ and $1 \times 10^{28}$\dunit\ at 100\tev, with a power-law scaling in rigidity with index 1/3, and different possible extents of the suppressed diffusion region. Particle injection starts at 60\,kyr after pulsar birth, with a high-energy spectral index assumed to have an average value of 2.4 and an efficiency set by default to 100\%. Using flatter injection indices does not profoundly modify the picture. For comparison, we show in the same plots the expected local positron flux if all nearby pulsars develop no halo, all particles being directly released in the \gls{ism}, diffusing out in average conditions for the Galactic plane without experiencing any kind of strong and extended confinement around the source. Injection efficiency is here again set to 100\% by default.

\begin{figure}[!t]
\begin{center}
\includegraphics[width=0.9\columnwidth]{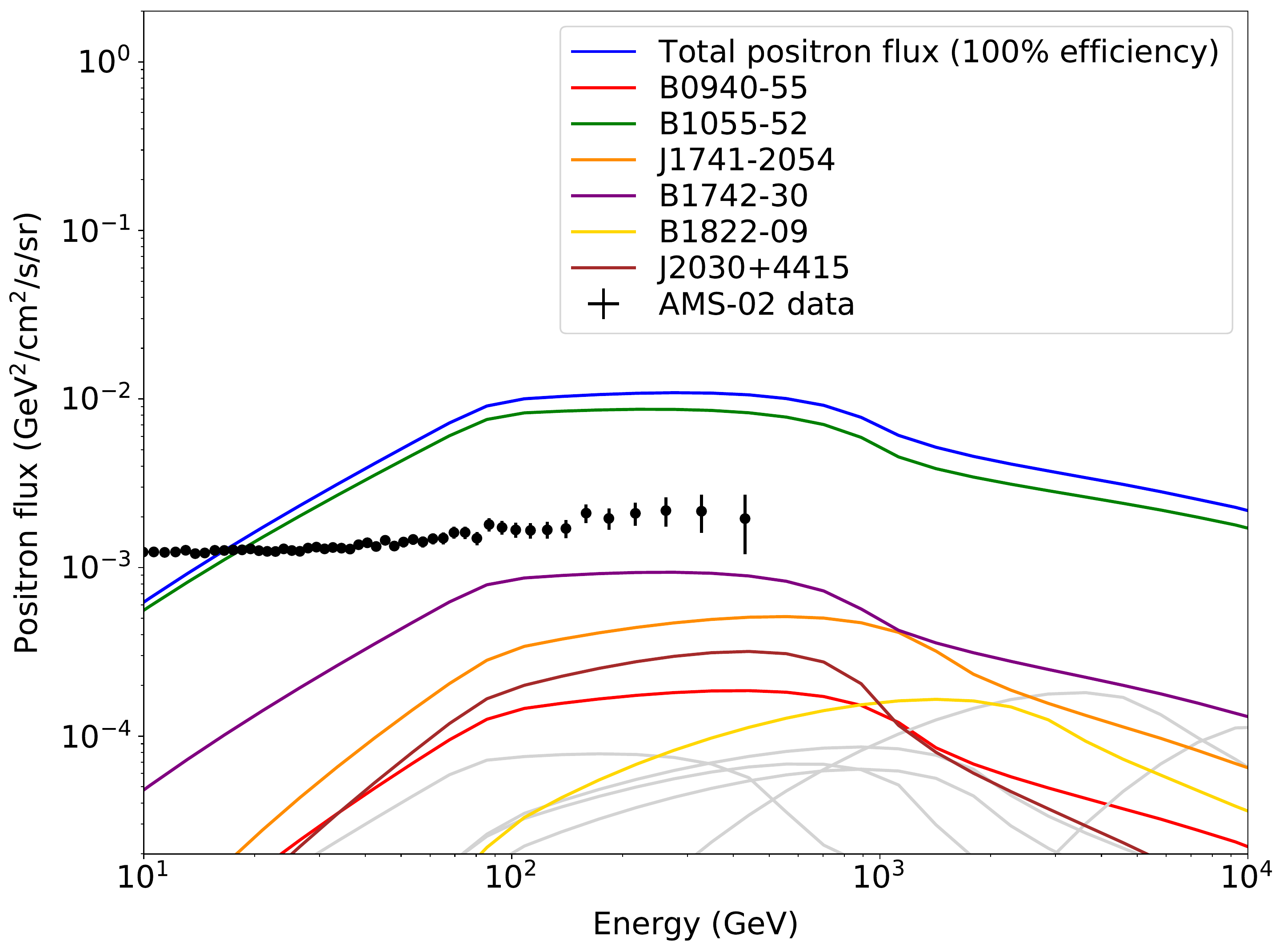}
\caption{Individual contributions to the total positron flux from known middle-aged pulsars within 1\,kpc, except J0633+1746 and B0656+14, assuming they develop a Geminga-like halo with a size of 50\,pc past an age of 60\,kyr. A default injection efficiency of 100\% is assumed for all pulsars, and only those contributing at the $\gtrsim 1$\% level or more are labeled.}
\label{fig:nearby:split}
\end{center}
\end{figure}

Figure \ref{fig:nearby:flux} shows that halos around nearby pulsars would yield a positron flux in excess of the AMS-02 measurement for average injection efficiencies $\gtrsim10-30$\%, depending on the extent and magnitude of diffusion suppression. Conversely, if positrons from nearby pulsars diffuse in the \gls{ism} without strong and extended confinement around the source, all known nearby pulsars taken together would saturate the local positron flux above $80-100$\gev for an average injection efficiency of $70-100$\%. It may seem counterintuitive that adding regions of inhibited diffusion around pulsars leads to lower injection efficiencies to match a given particle flux at a distant location. Such an effect was already discussed in \citet{Profumo:2018} in the case of Geminga. For particles suffering little from energy losses, typically $\lesssim1$\tev in the present context, halos actually slow down the release of particles and initially suppress their flux at a distant location. At later times, however, a high density of particles fills the entire suppressed diffusion region, and the diffusive flux escaping from its boundary is higher than what it would have been at the same time and position without a halo. For pulsars old enough to be in this stage and close enough to the observer, a smaller injection efficiency is thus required to match a given particle flux.

\begin{figure}[!t]
\begin{center}
\includegraphics[width=0.9\columnwidth]{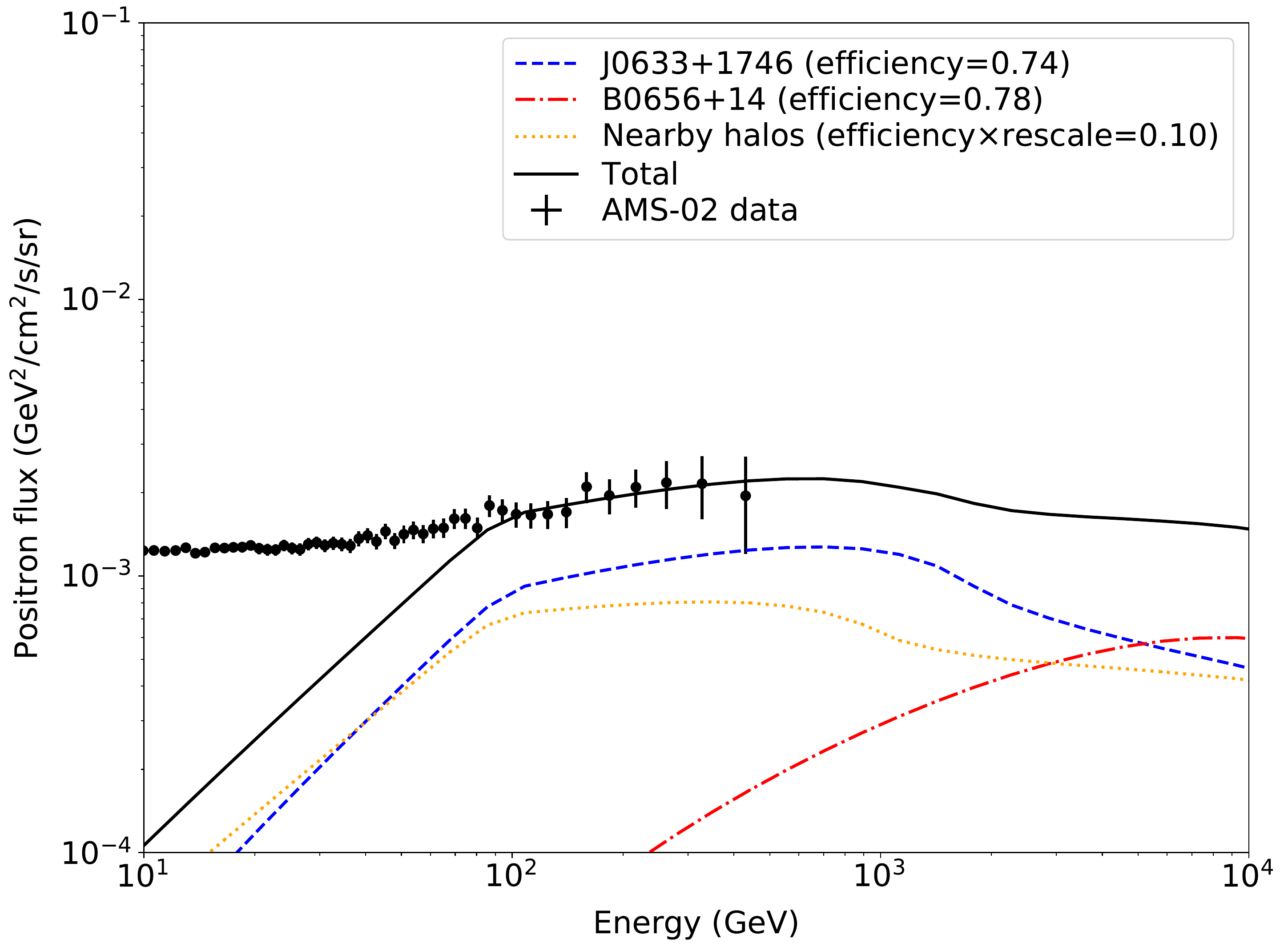}
\includegraphics[width=0.9\columnwidth]{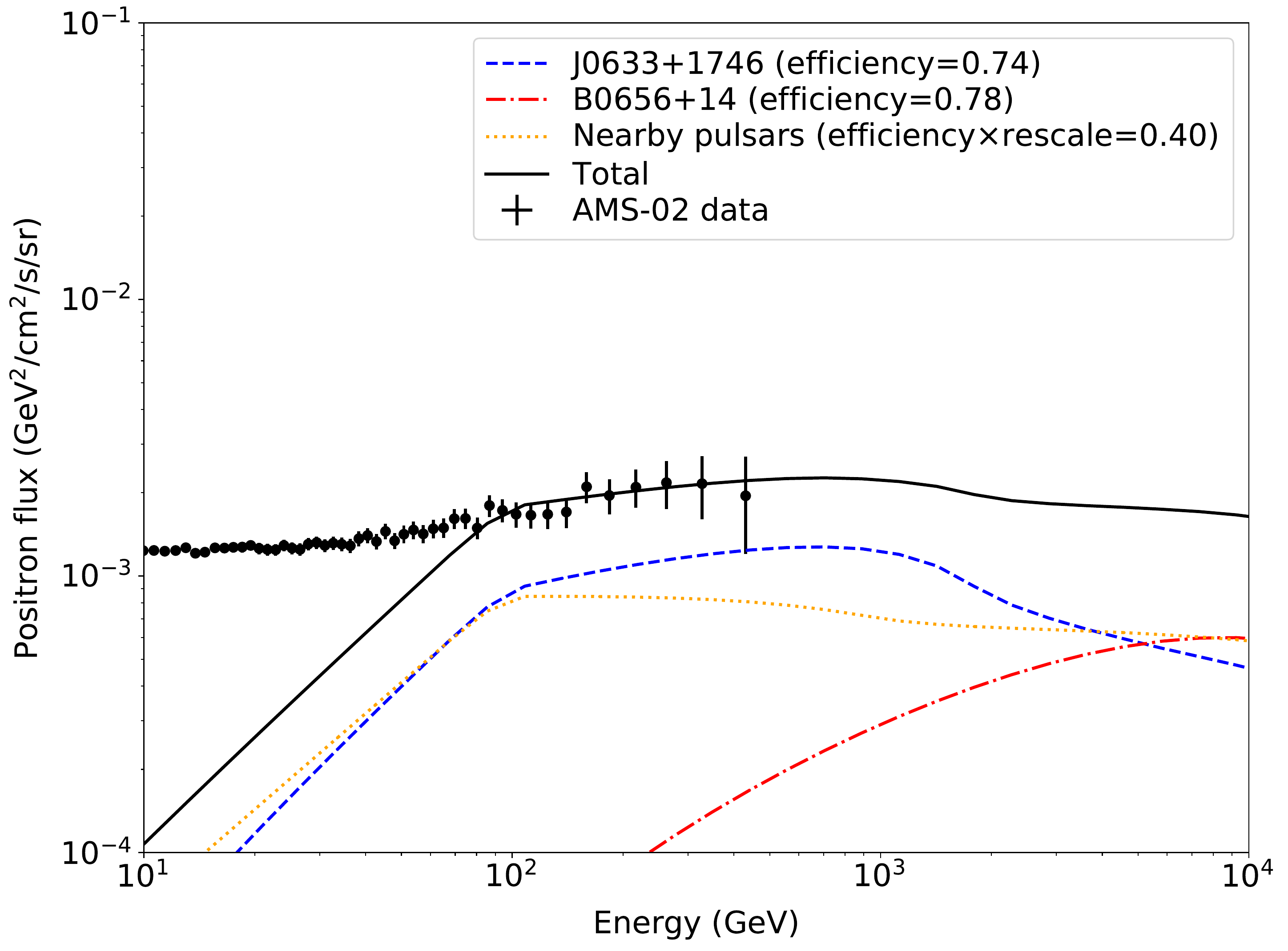}
\caption{Total positron flux from the halos around J0633+1746 and B0656+14 and the population of known nearby middle-aged pulsars under two hypotheses for the latter: all pulsars develop a halo (top panel) or none develop a halo and particles are released without confinement around the source (bottom panel). Suppressed diffusion regions of 30\,pc for both and an injection index of 2.4 were assumed for all halos. The injection efficiencies in J0633+1746 and B0656+14 were obtained by a fit to HAWC data, while the average value for other nearby pulsars was approximately set so that the sum of all contributions match the AMS-02 measurement in the $\sim0.1-1$\tev range.}
\label{fig:nearby:sum30pc}
\end{center}
\end{figure}

\begin{figure}[!t]
\begin{center}
\includegraphics[width=0.9\columnwidth]{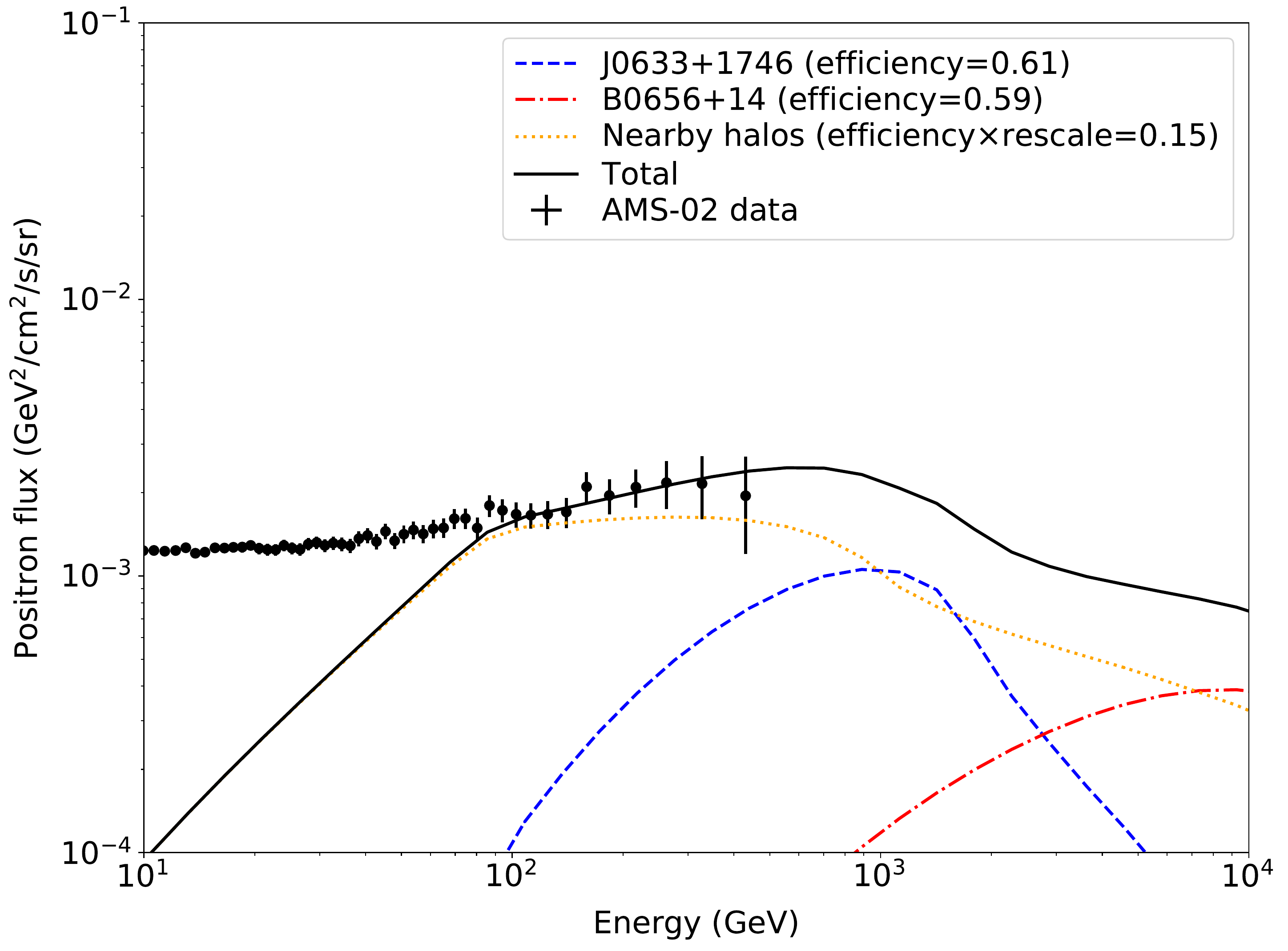}
\includegraphics[width=0.9\columnwidth]{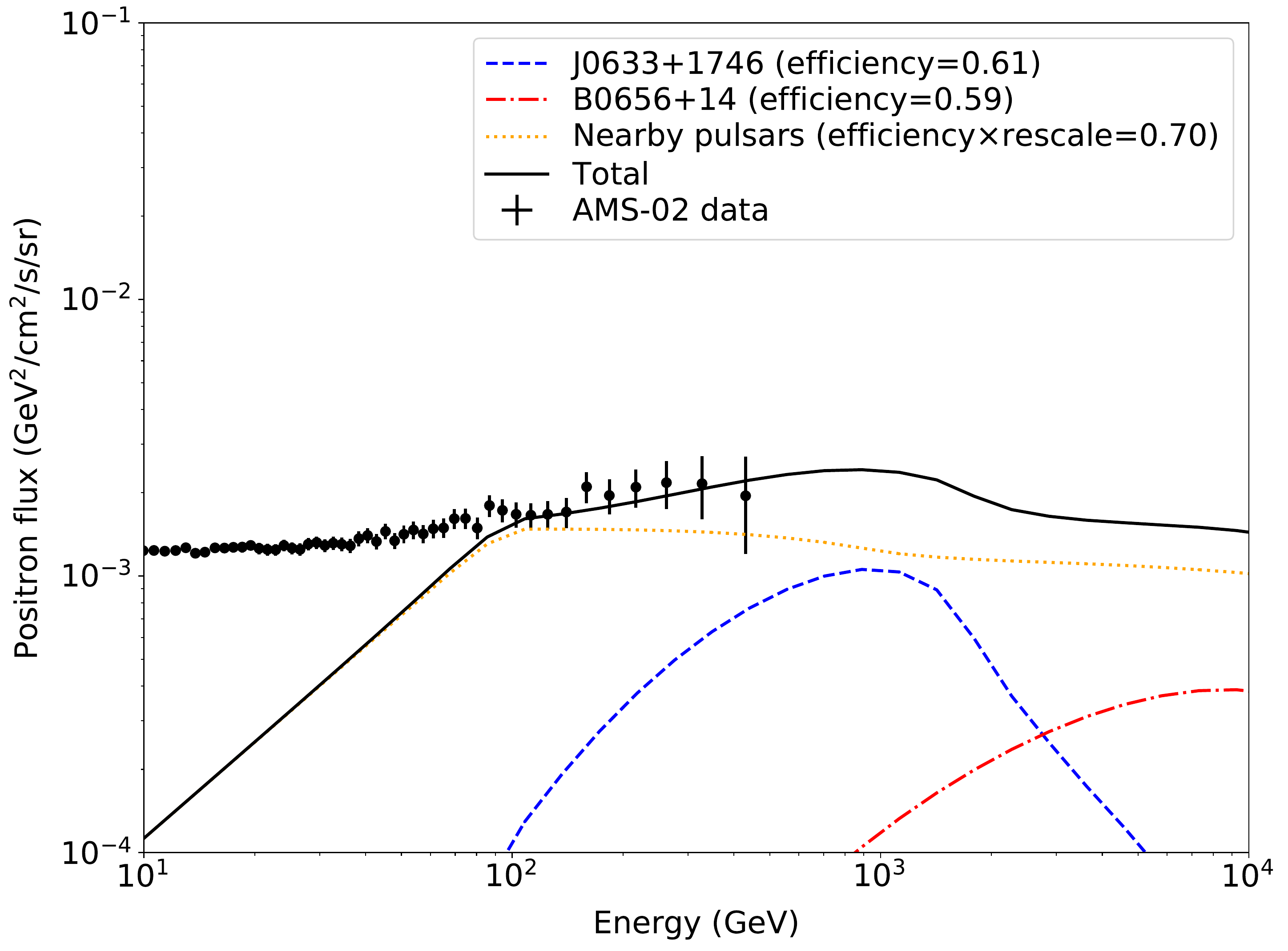}
\caption{Same as Fig. \ref{fig:nearby:sum30pc} for halos with a suppressed diffusion region of 80\,pc for Geminga and 50\,pc for others.}
\label{fig:nearby:sum50pc}
\end{center}
\end{figure}

The possible contribution from Geminga now needs to be folded in. According to the model setups listed in Sect. \ref{res:summary}, Geminga can be expected to contribute $30-50$\% or more of the positron flux over part or all of the $\sim0.1-1$\tev range. This leaves only about half or less of the positron flux to be accounted for from other pulsars. 
As above, we consider two extreme scenarios: all other pulsars develop halos (hereafter ``widespread halos scenario''), or none does (hereafter ``rare halos scenario''). This is illustrated in Figs. \ref{fig:nearby:sum30pc} and \ref{fig:nearby:sum50pc} for different halo model setups.

In the ``rare halos scenario'', the positron flux can be accounted for with all other known nearby pulsars having an average injection efficiency in the $40-70$\%, depending on the exact model for the halos around J0633+1746 and B0656+14. Such injection efficiencies are typical of young \gls{pwne} and similar to those for the halos around J0633+1746 and B0656+14. In that scenario, all pulsars have similar average properties in terms of injection spectrum and efficiency over the first few hundreds of kyr of their lifetime. Conversely, the ``widespread halos scenario'' involves small $10-15$\% injection efficiencies for all other nearby pulsars, at odds with the aforementioned values. This suggests that particle injection efficiency tends to decrease over the first few hundreds of kyr of a pulsar's lifetime, except in some cases like J0633+1746 and B0656+14 for some reason (which may explain why these two halos were detected first).

Quantitatively, however, the situation is more complicated. The typical injection efficiencies estimated above depend on the middle-aged population in our neighborhood, on the properties of the objects, and on the details of the formation of a suppressed diffusion region and injection of pairs into it. We discuss below those different features of the local pulsar population that may alter the above numbers and reduce the difference between the widespread and rare halos scenarios.

\textit{Distances}: As illustrated in Fig. \ref{fig:nearby:split}, a handful of nearby known objects dominate the positron flux, in particular PSR B1055-52. The estimated distance of 93\,pc for this pulsar, combined with its characteristic age of 535\,kyr and spin-down power of $3.01 \times 10^{34}$\punit, guarantee a dominant contribution to the local positron flux whatever the size of the suppressed diffusion region. In the case of a 120\,pc size, B1055-52 drives the total positron flux to very high values, well above those obtained for smaller sizes, because we then find ourselves inside its halo. The possible predominance of that object in the local positron flux was already noted in \citet{Fang:2019b}, while the scenario of a single pulsar within 100\,pc making up most of the positron flux above 100\gev (and most of the all-electron flux above 1\tev), was studied in more general terms in \citet{LopezCoto:2018b}. Assuming a distance of 350\,pc instead of 93\,pc for B1055-52, as suggested in \citet{Mignani:2010b}, the average injection efficiency needed in the ``widespread halos scenario'' to account for the observed positron flux increases by a factor $2-3$, while it increases by less than a factor 2 in the ``rare halos scenario''. This reduces the difference between both scenarios and brings the injection efficiencies for putative nearby halos closer to the typical expected range.

\textit{Ages}: Also illustrated in Fig. \ref{fig:nearby:split} is the fact that the local positron flux is dominated by several rather old objects with characteristic ages $\gtrsim500$\,kyr, like B1055-52, J2030+4415, or B1742-30. Whether halos can be developed around such old pulsars remains to be proven. They may have moved past the region of enhanced turbulence, or the latter may have damped, such that particles are not efficiently confined anymore. In that case, the average injection efficiency needed in the ``widespread halos scenario'' to account for the observed positron flux from all pulsars younger than 500\,kyr is increased, by nearly an order of magnitude, and becomes in agreement with the very high values inferred for young \gls{pwne}. Alternatively, the characteristic ages may be overestimating the true ages by a factor of a few, as already discussed in Sect. \ref{res:effpsr}. If those pulsars with characteristic ages $\gtrsim500$\,kyr actually have true ages in the $200-300$\,kyr range, the impact on the required injection efficiency is very limited, a decrease of the order of $10-20$\% at few hundreds of GeV energies, and the small injection efficiencies are overall preserved.

\textit{Completeness}: While the existence of very nearby objects like B1055-52 tends to lower the requirement on injection efficiency, so would the existence of more pulsars than considered here. The small sample we use may well underestimate the population of relevant objects because: (i) we considered only nearby pulsars within 1\,kpc, to simplify the calculations, while some contribution to the positron flux from objects up to $2-3$\,kpc can be expected just based on the typical diffusion scale length at 1\tev for average diffusion conditions; (ii) pulsars selected from the ATNF catalog were mostly identified in radio or X-rays through beamed emission periodically pointing toward our direction, and many more exist that we could not detect this way but would nevertheless contribute to the positron flux. Including all known middle-aged ATNF pulsars within 2\,kpc (41 objects instead of 14) yields an increase in the local positron flux by $\sim20$\% at most, and much less if B1055-52 actually is at a distance of 93\,pc and heavily dominates the flux. A more significant effect can be expected from unknown nearby pulsars. From estimates of the pulsar beaming fraction, the full population may count at least three times more objects than currently known \citep{Linden:2017}, maybe up to ten times. Including their contribution to the local positron flux would mechanically reduce by the same amount the average injection efficiencies needed to match the AMS-02 data, for both the ``widespread halos scenario'' and ``rare halos scenario'' although possibly in different proportions depending on the exact layout of objects in terms of distances and ages. It may well be, however, that the local population was probed to a relatively high degree of completeness because the proximity of pulsars favoured their detection in gamma-rays, especially since they are emitted with higher beaming fractions than pulsed radio signals. In \citet{Johnston:2020}, pulsars with spin-down power in the $10^{35}-10^{36}$ \punit, i.e. the decade above the one most relevant to us, have radio and gamma-ray beaming fractions of 0.33 and 0.50, respectively, with partial overlap, which suggests a correction for completeness by less than a factor of 2 for a sample of local objects that is not flux-limited, hence a limited correction on the above injection efficiency estimates.

\textit{Pair injection}: The conditions of particle injection also have an impact on the average injection efficiencies inferred in each scenario. While the high-energy spectral index $\alpha_2$ does influence the shape of the predicted contribution to the positron flux in the ``widespread halos scenario'', variations in the $2.0-2.4$ range have little effect on the average injection efficiency required to match the AMS-02 data because they preserve a maximum in the positron flux at $0.1-1$\tev energies and at about the same level in the E$^3$F(E) space. A more drastic effect results from assuming a single hard power-law injection spectrum with relatively low cutoff energy, as discussed in Sect. \ref{res:effinj}. A strong impact also arises from the injection start time. Replacing our default assumption of 60\,kyr by 20\,kyr leads to a predicted positron flux from all other nearby halos that is about twice as high, hence required injection efficiencies to match the AMS-02 measurement that are twice as low. More generally, any contribution to the injection from the early stages of a pulsar's evolution will push the required injection efficiency to match the AMS-02 data to smaller values, potentially very small because the pulsar was much more powerful at early times. In that respect, transitional objects between the classical young \gls{pwn} and evolved halo stages are potentially very important when it comes to their contribution to the local positron flux. Evaluating how the population of pairs injected in the early stages spreads out following the reverse-shock crushing of the original nebula and its mixing with the remnant's material would require a more involved modeling framework than the one used here for diffusive halos. To the best of our knowledge, this remains a poorly explored area.

The local population and the positron flux constraint therefore suggest different scenarios regarding the occurrence of pulsar halos:
\begin{enumerate}
\item Pulsar halos are rare: Locally, only J0633+1746 and B0656+14 have developed a halo, for some reason. The rest of the middle-aged pulsar population, known already to a high degree of completeness, contributes to the positron flux at Earth without strong confinement around the source. In this scenario, the properties of particle injection, spectrum and efficiency, are the same for all pulsars, from kyr-old objects powering \gls{pwne} to 100\,kyr-old objects, including J0633+1746 and B0656+14. This situation is preserved if the very nearby pulsar B1055-52 is located at a larger distance than 93\,pc and if, at the same time, the actual local pulsar population exceeds the known one by less than a factor 2.
\item Pulsar halos are common: Most middle-aged pulsars develop halos similar to those around J0633+1746 and B0656+14, but the average injection efficiency is of the order $10-15$\%, or less depending on the fraction of unknown local pulsars. This suggests a drop in injection efficiency by nearly one order of magnitude over the first few hundreds of kyr of a pulsar's lifetime, on average, with notable exceptions such as J0633+1746 and B0656+14. The required decrease in efficiency, however, can be reduced to just a factor of a few if the very nearby pulsar B1055-52 is located at a larger distance than 93\,pc, or even suppressed if pulsars with characteristic ages above 500\,kyr or so do not develop halos. On the other hand, the decrease in efficiency can be even stronger if a fraction of the particles produced at early times, when the pulsar was more powerful, manage to escape into the halo.
\end{enumerate}
Taking the population properties from the ATNF data base at face value, the first option may be favoured only for the sake of simplicity, as it requires no exceptional status for J0633+1746 and B0656+14. Yet, as discussed in the previous paragraphs, the exact properties of the population, especially the most nearby objects, are crucial because a handful of objects have a predominant contribution to the positron flux.

In the case where halos are common, very large halos do not seem favoured. As suggested in the top panel of Fig. \ref{fig:nearby:flux}, and this is even more marked in the absence of B1055-52, very large halos around nearby pulsars produce a rather peaked positron flux, with a maximum around $500-600$\gev for an average value of the injection index, much unlike the observed spectral shape, which is flat $\gtrsim100$\gev in the E$^3$F(E) space. The very existence of B1055-52, if it is indeed located very close to us, is an even stronger argument. It alone demonstrates that not all middle-aged pulsars develop very large halos with $\gtrsim100-120$\,pc sizes, otherwise this would place us within the halo of B1055-52, leading to a positron flux one to two orders of magnitude above the AMS-02 measurement for typical injection efficiencies of the order of a few tens of percent.

The hypothesis that halos could be relatively rare raises the question of their actual occurrence rate among pulsars. Based on the local sample of objects, assuming J0633+1746 and B0656+14 are the only halos, the occurrence rate would lie in the $\sim5-10$\% range depending on the fraction of the local pulsar population that we can assume as known: typically $\sim12-13$\% if the 16 known pulsars within 1\,kpc are all those that exist, and half of that if there are twice as many. We note that rarity should probably be understood in a broad sense: either halos are long-lived structures that occur rarely among the pulsar population (the option we most directly test here with our static model), or they are an infrequent evolutionary phase in a larger number of pulsars. Both possibilities could manifest in the same way in terms of contribution to the local positron flux, depending on the parameters of the problem.

While a number of extended TeV sources are positionally coincident with middle-aged pulsars, for instance the list presented in \citet{Linden:2017} and updated in \citet{Albert:2020}, formal identification is lacking in most cases such that the idea that halos are generally rare among middle-aged pulsars is perfectly viable. In the list of candidate halos from \cite{Albert:2020}, only 8 out of 16 middle-aged pulsars observable with HAWC can be associated to a 3HWC source, despite a very loose criterion of positional coincidence within 1\deg, and alternative interpretations are available for many of these. For instance, the source associated to PSR J1913+1011 may well be an \gls{snr} \citep[see][]{Zhang:2020b}, while the source associated to PSR J2032+4127 is most likely a classical \gls{pwn} associated to the gamma-ray binary \citep[see][]{Aliu:2014,Lyne:2015,AlbaceteColombo:2020}. In several other cases, the associations involve powerful pulsars with spin-down power $\gtrsim10^{36}$\punit\ and characteristic ages $\sim$100\,kyr, like PSR J1831-0952 and PSR J1925+1720, such that a classical \gls{pwn} is the most likely origin for the gamma-ray emission, as suggested in more general terms in \citet{Giacinti:2020}. Overall, from the list in \citet{Albert:2020}, only 3 out of 16 middle-aged pulsars are associated to sources that can be considered as solid halo candidates (J0633+1746, B0656+14, and B0540+23), eventually yielding an occurrence rate similar to the one provided above, from a different approach.

The above results suggest a range of possibilities for the composition of the positron flux above $80-100$\gev: (i) if the halos around J0633+1746 and B0656+14 have large extents of 80-100\,pc, and the corresponding positron flux is heavily suppressed, the spectrum is dominated by particles freely escaping from nearby middle-aged pulsars and falls off more or less abruptly in the TeV range depending on the actual injection spectrum in nearby pulsars (e.g. spectral index in the $2.4-2.8$ range, existence of a cutoff at lower energies than assumed here; see Fig. \ref{fig:nearby:sum50pc}); (ii) if the halos around J0633+1746 and B0656+14 have moderate extents of $30-50$\,pc, their summed contribution to the positron flux dominates above 100\gev, with a flat or slightly rising spectrum $\gtrsim1$\tev \citep[see Fig. \ref{fig:nearby:sum30pc} and also][]{Fang:2019b}. The actual trend may hopefully be revealed in future releases of AMS-02 data.

Last, we emphasize that the above results and conclusions hold in the adopted model framework. It is perfectly conceivable that the suppressed diffusion region has both time and energy-dependent properties, e.g. extent or onset of diffusion suppression depending on particle energy. Such solutions would certainly help to relax the local positron flux constraint and thus allow for halos to be more widespread (at the expense, of course, of more free parameters).

% Conclusions
\section{Conclusions}
\label{conclu}

The discovery of extended gamma-ray emission structures around a growing but still small number of middle-aged pulsars, interpreted as halos formed by the inverse-Compton scattering of ambient photon fields by electron-positron pairs, has raised the question of their commonness in the Galaxy. Observations of the nearby PSR J0633+1746 and PSR B0656+14 almost directly indicate diffusion suppression by factors reaching $2-3$ orders of magnitude at $\sim100$\tev, and complementary analyses have suggested that this inhibited diffusion could extend over at least $100-120$\,pc scales. These numbers, however, seem extreme when compared to what can be achieved in recent theoretical attempts to account for the phenomenon, under reasonable and representative assumptions for particle acceleration and release in these systems.

Using the phenomenological framework of a static two-zone diffusion model, we searched for parameter setups minimizing as much as possible the extent and magnitude of pair confinement around J0633+1746 and B0656+14. We compared model predictions to data from {\it Fermi}-LAT, HAWC, and AMS-02, under the requirement that non-thermal particles released in the halos past an age of 60\,kyr are representative of \gls{pwne} in terms of injection spectrum and efficiency.  

In this framework, the magnitude of diffusion suppression at 100\tev particle energies, with respect to average conditions in the Galactic plane, appears tightly constrained for both objects. The case of J0633+1746 is the most extreme, with diffusion suppression at a level of $\sim500$ constrained by both the HAWC and AMS-02 observations and the total energetics of the pulsar. The age and distance of J0633+1746 are such that a suppressed diffusion region of moderate extent $\lesssim50$\,pc yields an excessive contribution to the local positron flux and requires an excessive injection efficiency for commonly assumed parameters of the problem. These issues can be circumvented if the Geminga pulsar has a true age $\lesssim200$\,kyr, smaller than its characteristic age, and/or if the injection spectrum is relatively flat, which implies a flux in the {\it Fermi}-LAT range below the level of the claimed detection. In these cases, solutions with suppressed diffusion regions as small as 30\,pc can be found. The case of B0656+14, with diffusion suppression at a level of $\sim100$, is less constrained, owing to the non-detection in the {\it Fermi}-LAT range and to the smaller age and larger distance of the pulsar compared to Geminga. This translates into a much smaller contribution to the local positron flux, peaking at energies above the current reach of AMS-02, which makes it easier to fit within the positron flux constraint for a wide range of suppressed diffusion region sizes. For both pulsars, depending on the exact model setup, injection efficiencies above a few tens of percent are required, similar to the values inferred for the younger \gls{pwne}. In these model setups, the contribution of both pulsar halos to the local positron flux is significant and could even dominate above a few hundreds of GeV. In particular, Geminga would account for about $30-50$\% or more of the positron flux over part or all of the $\sim0.1-1$\tev range.

If all other nearby middle-aged pulsars develop halos similar to our minimal model setups for J0633+1746 and B0656+14, their combined positron flux complemented by the contribution from Geminga would saturate the AMS-02 measurement above $\gtrsim100$\gev for an average injection efficiency that is much smaller than the values inferred for the halos around J0633+1746 and B0656+14 or with the values inferred for younger \gls{pwne}. In that scenario, extremely large suppressed diffusion regions of $100-120$\,pc or more, are disfavoured for the following reasons: (i) this would place us inside the halo of PSR B1055-52 with, as a consequence of its proximity and age, an extremely large positron flux overwhelming the AMS-02 measurement for a typical injection efficiency; (ii) this raises the question of what exactly contributes to the local $\gtrsim100$\gev positron flux since the cumulative contribution from very large halos is strongly peaked at $0.5-1$\tev, much unlike the flat spectral shape observed with AMS-02; (iii) this remains in tension with what seems achievable in most theoretical models.

Conversely, if positrons from other nearby pulsars are released in the \gls{ism} without any kind of confinement around the source, their local flux complements the contribution from Geminga for average injection efficiencies of a few tens of percent for all pulsars, from kyr-old objects powering \gls{pwne} to 100\,kyr-old objects like J0633+1746 and B0656+14. It thus seems a simpler scenario to consider that most middle-aged pulsars do not develop halos similar to those around J0633+1746 and B0656+14, or even no halo at all. Yet, the quantitative difference between the rare or common halos scenarios, in terms of required injection efficiency to fit the positron flux constraint, depends a lot on the actual properties of the local population of pulsars and on the still uncertain physics driving the formation and evolution of halos.

Based on local objects, the occurrence rate of halos among middle-aged pulsars could be as low as $\sim5-10$\%. The local positron flux at $\sim0.1-1$\tev energies would therefore be attributed to a few dozens nearby middle-aged pulsars rapidly releasing pairs into the \gls{ism}, with a possible contribution over part or most of the range by J0633+1746, and at higher energies by B0656+14, depending on their properties. Positron flux measurements in the TeV range would therefore be instrumental in constraining the size of the halos around J0633+1746 and B0656+14, and the possible existence of other instances of the phenomenon in our neighborhood.

\begin{acknowledgements}
The authors acknowledge financial support by CNES for the exploitation of {\it Fermi}-LAT observations, and by ANR for support to the GAMALO project under reference ANR-19-CE31-0014. We thank R\'uben L\'opez-Coto for providing the HAWC intensity data in a convenient format, Carmelo Evoli for insightful discussions and comments on the manuscript, and Benedikt Schroer for spotting a mistake in an initial version of the results. We also thank the anonymous referee for very valuable comments that contributed to improve the quality of the manuscript. This work has made use of the SIMBAD database, operated at CDS, Strasbourg, France, and of NASA's Astrophysics Data System Bibliographic Services. The preparation of the figures has made use of the following open-access software tools: Astropy \citep{Astropy:2013}, Matplotlib \citep{Hunter:2007}, NumPy \citep{VanDerWalt:2011}, and SciPy \citep{Virtanen:2020}.
\end{acknowledgements}

\bibliographystyle{aa}
\bibliography{Biblio/Halo.bib,Biblio/ISM.bib,Biblio/ISRF.bib,Biblio/Pulsars.bib,Biblio/DataAnalysis.bib,Biblio/Physics.bib,Biblio/Fermi.bib,Biblio/CosmicRayEscape.bib,Biblio/CosmicRayAcceleration.bib,Biblio/CosmicRayTransport.bib,Biblio/CosmicRayMeasurements.bib,Biblio/GalacticDiffuseEmission.bib,Biblio/SNobservations.bib,Biblio/SNRobservations,Biblio/LMC.bib}

\end{document}